\documentclass[english]{elsarticle}
\usepackage[T1]{fontenc}
\usepackage[latin9]{inputenc}
\usepackage{geometry}
\usepackage{array}
\usepackage{booktabs}
\usepackage{bm}
\usepackage{amsmath}
\usepackage{graphicx}
\usepackage{esint}

\makeatletter

\providecommand{\tabularnewline}{\\}

\newcommand{\apjl}{ApJL}
\newcommand{\apj}{ApJ}
\newcommand{\apjs}{ApJS}
\newcommand{\jgr}{J. Geophys. Res.}
\newcommand{\grl}{Geophys. Res. Letter}
\newcommand{\aap}{A~\&~A}
\newcommand{\aapr}{Astron.~\&~Astrophys.~Rev.}
\newcommand{\mnras}{MNRAS}
\newcommand{\solphys}{Sol.Phys.}
\newcommand{\physrep}{Phys. Rep.}

\usepackage{babel}

\makeatother

\usepackage{babel}
\begin{document}
\begin{frontmatter}{}

\title{Nonkinematic solar dynamo models with double-cell meridional circulation }

\author{V.V. Pipin}

\address{Institute of Solar-Terrestrial Physics, Russian Academy of Sciences,
Irkutsk, 664033, Russia}
\begin{abstract}
Employing the standard solar interior model as input we construct
a dynamically-consistent nonlinear dynamo model that takes into account
the detailed description of the $\Lambda$- effect, turbulent pumping,
magnetic helicity balance, and magnetic feedback on the differential
rotation and meridional circulation. The background mean-field hydrodynamic
model of the solar convection zone accounts the solar-like angular
velocity profile and the double-cell meridional circulation. We investigate
an impact of the nonlinear magnetic field generation effects on the
long-term variability and properties of the magnetic cycle. The nonlinear
dynamo solutions are studied in the wide interval of the $\alpha$
effect parameter from a slightly subcritical to supercritical values.
It is found that the magnetic cycle period decreases with the increasing
cycle's magnitude. The periodic long-term variations of the magnetic
cycle are excited in case of the overcritical $\alpha$ effect. These
variations result from the hemispheric magnetic helicity exchange.
It depends on the magnetic diffusivity parameter and the magnetic
helicity production rate. The large-scale magnetic activity modifies
the distribution of the differential rotation and meridional circulation
inside convection zone. It is found that the magnetic feedback on
the global flow affects the properties of the long-term magnetic cycles.
We confront our findings with solar and stellar magnetic activity
observations. 
\end{abstract}
\end{frontmatter}{}

\section{Introduction}

There is a commonly accepted idea that the sunspot activity is produced
by the large-scale toroidal magnetic field which is generated inside
the convection zone by means of the differential rotation \citep{P55}.
The theory explains the 11-year solar cycle as a result of the large-scale
dynamo operating in the solar interior, where, in addition to the
magnetic fields generated by the differential rotation, the helical
convective motions transforms the energy of the toroidal magnetic
fields to poloidal. The effect of meridional circulation on the large-scale
dynamo is not well understood. It is the essential part of the flux-transport
dynamo model scenario \citep{choud95,cd99} to explain the equatorward
drift of the toroidal magnetic field in the solar cycle. Here, it
assumed that the toroidal field at the bottom of the convection zone
forms sunspot activity. Feasibility of this idea can be questioned
both the observational and theoretical arguments \citep{b05}. The
distributed dynamo models can be constructed with \citep{2002AA...390..673B,2008AA...483..949J,2014AA563A18P}
and without \citep{moss00M,pip13M} effect of meridional transport
of the large-scale magnetic field.

Recent results of helioseismology reveal the double-cell meridional
circulation structure \citep{Zhao13m,2017ApJ845.2B}. It demolishes
the previously accepted scenario of the flux-transport models \citep{2014ApJ782.93H,2016MNRAS.456.2654W,2016ApJ832.9H}.
Contrary, \citet{PK13} showed the distributed dynamo models can reproduce
observations with regards to the subsurface rotational shear layer
and the double-cell meridional circulation. In their model, the double-cell
meridional circulation was modeled in following to results of helioseismology
of \citet{Zhao13m}. The effect of the multi-cell meridional circulation
on the global dynamo was also studied in the direct numerical simulations
\citep{kap2012,2016ApJ819.104G,2018AA609A..51W}. There were no attempts
to construct the non-kinematic mean-field dynamo models with regards
to the multi-cell meridional circulation.

The standard mean-field models of the solar differential rotation
predict a one-cell meridional circulation per hemisphere. This contradicts
to the helioseismology inversions and results of direct numerical
simulations. In the mean-field theory framework, the differential
rotation of the Sun is explained as a result of the angular momentum
transport by the helical convective motions. Similarly to a contribution
of the $\alpha$ effect in the mean-electromotive force, i.e., 
\[
\mathbf{\mathcal{E}}=\left\langle \mathbf{u}\times\mathbf{b}\right\rangle =\hat{\alpha}\circ\left\langle \mathbf{B}\right\rangle +\dots,
\]
where $\mathbf{u}$ is the turbulent velocity $\mathbf{u}$, and $\mathbf{b}$
is the turbulent magnetic field, the $\Lambda$-effect, (e.g., \citealp{1989drsc.book.....R})
appears as the non-dissipative part of turbulent stresses\textbf{
\[
\hat{T}_{ij}=\left\langle u_{i}u_{j}\right\rangle =\Lambda_{ijk}\Omega_{k}+\dots
\]
}where $\boldsymbol{\Omega}$ is the angular velocity. The structure
of the meridional circulation is determined by directions of the non-diffusive
angular momentum transport due to the $\Lambda$ effect \citep{2017ApJ835.9B}.
In particular, the vertical structure of the meridional circulation
depends on the sign of the radial effect. It was found that the double-cell
meridional circulation can be explained if the of $\Lambda$-effect
changes sign in the depth of the convection zone. \citet{2018ApJ854.67P}
showed that this effect can result from the radial inhomogeneity of
the convective turnover timescale. It was demonstrated that if this
effect is taken into account then the solar-like differential rotation
and the double-cell meridional circulation are both reproduced by
the mean-field model .

In this paper, we apply the meridional circulation profile, which
is calculated from the solution of the angular momentum balance to
the nonkinematic dynamo models. Previously, the similar approach was
applied by \citet{1992AA...265..328B} and \citet{2006ApJ...647..662R}
in the distributed and the flux-transport models with one meridional
circulation cell as the basic stage in the non-magnetic case.

Our main goal is to study how the double-cell meridional circulation
affects the nonlinear dynamo generation of the large-scale magnetic
field. The magnetic feedback on the global flow can result in numerous
physical phenomena such as the torsional oscillations \citep{1982SoPh...75..161L,2011JPhCS271a2074H},
the long-term variability of the magnetic activity \citep{sok1994AA,2014JGRA119.6027F}
etc. The properties of the nonlinear evolution depend on the dynamo
governing parameters such as amplitude of turbulent generation of
the magnetic field by the $\alpha$ effect, as well as the other nonlinear
processes involved in the dynamo, i.e., the dynamo quenching by the
magnetic buoyancy effect \citep{kp93,tob98} and the magnetic helicity
conservation \citep{kleruz82}. We study if the long-term variation
of magnetic activity can result from the increasing level of turbulent
generation of magnetic field by the $\alpha$ effect. The increasing
of the $\alpha$ effect results to an increase of the magnetic helicity
production. This affects the large-scale magnetic field generation
by means of the magnetic helicity conservation. Hence, the magnetic
helicity balance has to be taken into account. 

{It is hardly possible to consider in full all the goals within
one paper. From our point of view, the most important tasks includes:
construction of the solar-type dynamo model with the multi-cell meridional
circulation and studying the principal nonlinear dynamo effects. The
latter includes the magnetic helicity conservation and the nonkinematic
effects due to the magnetic feedback on the large-scale flow. Accordingly,
the paper is organized as follows.} Next Section describes the hydrodynamic,
thermodynamic and magnetohydrodynamic parts of the model. Then, I
present an attempt to construct the solar-type dynamo model and discuss
the effect of the turbulent pumping on the properties of the dynamo
solution. {The next subsections consider result for the principal
nonlinear dynamo effects. They deals with the global flows variations,
the Grand activity cycles and the magnetic cycle variations of the
thermodynamic parameters in the model.} The paper is concluded with
a discussion of the main results using results of other theoretical
studies and results of observations.

\section{Basic equations.}

\subsection{The angular momentum balance\label{subsec:am}}

We consider the evolution of the axisymmetric large-scale flow, which
is decomposed into poloidal and toroidal components: $\mathbf{\overline{U}}=\mathbf{\overline{U}}^{m}+r\sin\theta\Omega\hat{\mathbf{\boldsymbol{\phi}}}$,
where $\boldsymbol{\hat{\phi}}$ is the unit vector in the azimuthal
direction. The mean flow satisfies the stationary continuity equation,
\begin{equation}
\boldsymbol{\nabla}\cdot\overline{\rho}\mathbf{\overline{U}}=0,\label{eq:cont}
\end{equation}
Distribution of the angular velocity inside convection zone is determined
by conservation of the angular momentum \citep{1989drsc.book.....R}:
\begin{eqnarray}
\frac{\partial}{\partial t}\overline{\rho}r^{2}\sin^{2}\theta\Omega & = & -\boldsymbol{\nabla\cdot}\left(r\sin\theta\overline{\rho}\left(\hat{\mathbf{T}}_{\phi}+r\sin\theta\Omega\mathbf{\overline{U}^{m}}\right)\right)\label{eq:angm}\\
 & + & \boldsymbol{\nabla\cdot}\left(r\sin\theta\frac{\overline{\mathbf{B}}\overline{B}_{\phi}}{4\pi}\right).\nonumber 
\end{eqnarray}
To determine the meridional circulation we consider the azimuthal
component of the large-scale vorticity , $\omega=\left(\boldsymbol{\nabla}\times\overline{\mathbf{U}}^{m}\right)_{\phi}$
, which is governed by equation: 
\begin{eqnarray}
\frac{\partial\omega}{\partial t}\!\!\! & \negthinspace\!=\!\!\!\! & r\sin\theta\boldsymbol{\nabla}\cdot\left(\frac{\hat{\boldsymbol{\phi}}\times\boldsymbol{\nabla\cdot}\overline{\rho}\hat{\mathbf{T}}}{r\overline{\rho}\sin\theta}-\frac{\mathbf{\overline{U}}^{m}\omega}{r\sin\theta}\right)\!\!+r\sin\theta\frac{\partial\Omega^{2}}{\partial z}\label{eq:vort}\\
 & +\!\!\! & \frac{1}{\overline{\rho}^{2}}\left[\boldsymbol{\nabla}\overline{\rho}\times\boldsymbol{\nabla}\overline{p}\right]_{\phi}\!\!\nonumber \\
 & + & \!\frac{1}{\overline{\rho}^{2}}\left[\!\!\boldsymbol{\nabla}\overline{\rho}\times\left(\!\!\boldsymbol{\nabla}\frac{\overline{\mathbf{B}}^{2}}{8\pi}-\frac{\left(\overline{\mathbf{B}}\boldsymbol{\cdot\nabla}\right)\overline{\mathbf{B}}}{4\pi}\!\right)\!\!\right]_{\phi},\nonumber 
\end{eqnarray}
The turbulent stresses tensor, $\hat{\mathbf{T}}$, is written in
terms of small-scale fluctuations of velocity and magnetic field:
\begin{equation}
\hat{T}_{ij}=\left(\left\langle u_{i}u_{j}\right\rangle -\frac{1}{4\pi\overline{\rho}}\left(\left\langle b_{i}b_{j}\right\rangle -\frac{1}{2}\delta_{ij}\left\langle \mathbf{b}^{2}\right\rangle \right)\right).\label{eq:stres}
\end{equation}
where ${\partial/\partial z=\cos\theta\partial/\partial r-\sin\theta/r\cdot\partial/\partial\theta}$
is the gradient along the axis of rotation. The turbulent stresses
affect generation and dissipation of large-scale flows, and they are
affected by the global rotation and magnetic field. The magnitude
of the kinetic coefficients in tensor $\hat{\mathbf{T}}$ depends
on the rms of the convective velocity, ${u}'$, the strength of the
Coriolis force and the strength of the large-scale magnetic field.
The effect of the Coriolis force is determined by parameter $\Omega^{*}=2\Omega_{0}\tau_{c}$,
where $\Omega_{0}=2.9\times10^{-6}$rad/s is the solar rotation rate
and $\tau_{c}$ is the convective turnover time. The effect of the
large-scale magnetic field on the convective turbulence is determined
by parameter $\beta=\left\langle \left|\mathbf{B}\right|\right\rangle /\sqrt{4\pi\overline{\rho}u'^{2}}$.

The magnetic feedback on the coefficients of turbulent stress tensor
$\hat{\mathbf{T}}$ was studied previously with the mean-field magnetohydrodynamic
framework \citep{rob-saw}. In our model we apply analytical results
of \citet{1994AN....315..157K}, \citet{kit-rud-kuk} and \citet{kuetal96}.
The analytical expression for $\hat{\mathbf{T}}$ is given in Appendix.
It was found that the standard components of the nondissipative of
$\hat{\mathbf{T}}$ ($\Lambda$-effect) are quenched with the increase
of the magnetic field strength as $\beta^{-2}$ and the magnetic quenching
of the viscous parts is the order of $\beta^{-1}$. Also, there is a non-trivial
effect inducing the latitudinal angular momentum flux proportional
to the magnetic energy \citep{kit-rud-kuk,kuetal96}. This effect
is quenched as $\beta^{-2}$ for the case of the strong magnetic field.
Implications of the magnetic feedback on the turbulent stress tensor
$\hat{\mathbf{T}}$ were discussed in the models of solar torsional
oscillations and Grand activity cycles \citep{kit-rud-kuk,kuetal96,p99,1999AA343.977K}.
The analytical results of the mean-field theory are in qualitative
agreement with the direct numerical simulations \citep{2007AN....328.1006K,kap2011,2017arXiv171208045K}.

Profile of $\tau_{c}$ (as well as profiles of $\overline{\rho}$
and other thermodynamic parameters) is obtained from a standard solar
interior model calculated using the MESA code \citep{mesa11,mesa13}.
The rms velocity, $u'$, is determined in the mixing length approximations
from the gradient of the mean entropy, $\overline{s}$, 
\begin{equation}
u'=\frac{\ell}{2}\sqrt{-\frac{g}{2c_{p}}\frac{\partial\overline{s}}{\partial r}},\label{eq:uc}
\end{equation}
where $\ell=\alpha_{MLT}H_{p}$ is the mixing length, $\alpha_{MLT}=2.2$
is the mixing length theory parameter, and $H_{p}$ is the pressure
scale height. For a non-rotating star the ${u}'$ profile corresponds
to results of the MESA code. The mean-field equation for heat transport
takes into account effects of rotation and magnetic field \citep{2000ARep...44..771P}:
\begin{equation}
\overline{\rho}\overline{T}\left(\frac{\partial\overline{s}}{\partial t}+\left(\overline{\mathbf{U}}\cdot\boldsymbol{\nabla}\right)\overline{s}\right)=-\boldsymbol{\nabla}\cdot\left(\mathbf{F}^{conv}+\mathbf{F}^{rad}\right)-\hat{T}_{ij}\frac{\partial\overline{U}_{i}}{\partial r_{j}}-\frac{1}{4\pi}\boldsymbol{\mathcal{E}}\cdot\nabla\times\boldsymbol{\overline{B}},\label{eq:heat}
\end{equation}
where, $\overline{\rho}$ and $\overline{T}$ are the mean density
and temperature, $\boldsymbol{\mathcal{E}}=\left\langle \mathbf{u\times b}\right\rangle $
is the mean electromotive force. The Eq.(\ref{eq:heat}) includes
the thermal energy loss and gain due to generation and dissipation
of large-scale flows. The last term of the Eq.(\ref{eq:heat}) takes
into account effect of thermal energy exchange because of dissipation
and generation of magnetic field \citep{2000ARep...44..771P}. In
derivation of the mean-field heat transport equation (see, \citealp{2000ARep...44..771P}),
it was assumed that the magnetic and rotational perturbations of the
reference thermodynamic state are small. Also the parameters of the
reference state are given independently by the MESA code.

For the anisotropic convective flux we employ the expression suggested
by \citet{1994AN....315..157K} (hereafter KPR94), 
\begin{equation}
F_{i}^{conv}=-\overline{\rho}\overline{T}\chi_{ij}\nabla_{j}\overline{s}.\label{conv}
\end{equation}
The further details about dependence of the eddy conductivity tensor
$\chi_{ij}$ from effects of both the global rotation and large-scale
magnetic field are given in Appendix. The diffusive heat transport
by radiation reads, 
\[
\mathbf{F}^{rad}=-c_{p}\overline{\rho}\chi_{D}\boldsymbol{\nabla}T,
\]
where 
\[
\chi_{D}=\frac{16\sigma\overline{T}^{3}}{3\kappa\overline{\rho}^{2}c_{p}}.
\]
Both the eddy conductivity and viscosity are determined from the mixing-length
approximation: 
\begin{eqnarray}
\chi_{T} & = & \frac{\ell^{2}}{4}\sqrt{-\frac{g}{2c_{p}}\frac{\partial\overline{s}}{\partial r}},\label{eq:chi}\\
\nu_{T} & = & \mathrm{Pr_{T}}\chi_{T},\label{eq:nu}
\end{eqnarray}
where $\mathrm{Pr_{T}}$ is the turbulent Prandtl number. {Note,
that in Eq\eqref{eq:chi} we employ factor $1/2$ instead of $1/3$.
With this choice the distribution of the mean entropy gradient, which
results from solution of the Eq\eqref{eq:heat} for the nonrotating
and nonmagnetic case is close to results of the MESA code.} It is
assumed that $\mathrm{Pr_{T}}={\displaystyle 3/4}$. {This
corresponds to the theoretical results of KPR94}. For this choice
we have the good agreement with solar angular velocity latitudinal
profile. We assume that the solar rotation rate corresponds to rotation
rate of solar tachocline at 30$^{\circ}$ latitude, i.e., $\Omega_{0}/2\pi=430$nHz
\citep{1997SoPh170.43K}. We employ the stress-free boundary conditions
in the hydrodynamic part of the problem. For the Eq(\ref{eq:heat})
the thermal flux at the bottom is taken from the MESA code. At the
top, the thermal flux from the surface is approximated by the flux
from a blackbody: 
\begin{equation}
F_{r}=\frac{L_{\odot}}{4\pi r^{2}}\left(1+4\frac{T_{e}}{T_{eff}}\frac{\overline{s}}{c_{p}}\right),\label{eq:flx}
\end{equation}
where where $T_{eff}$ is the effective temperature of the photosphere
and $T_{e}$ is the temperature at the outer boundary of the integration
domain.

\begin{figure}
\includegraphics[width=0.8\columnwidth]{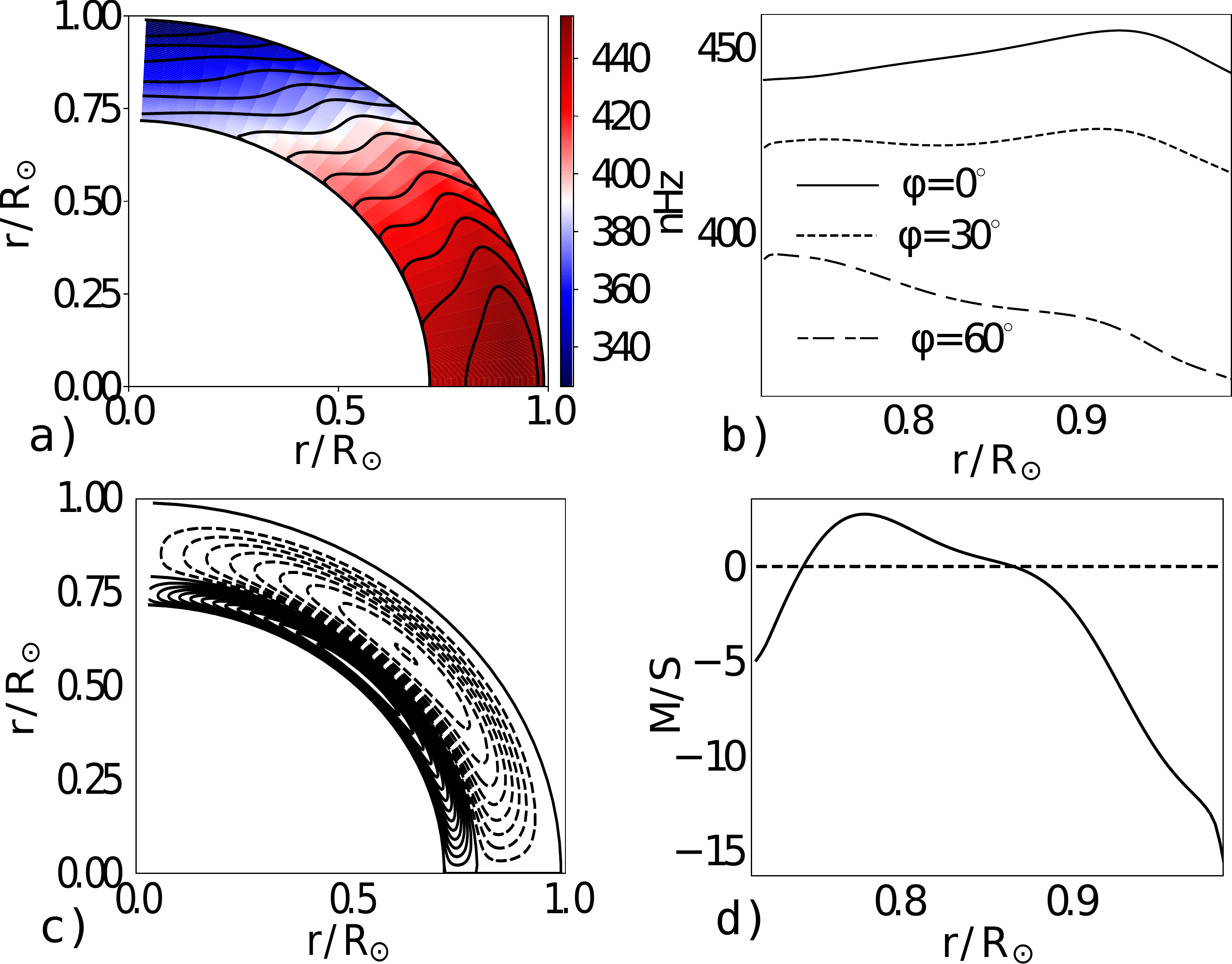}

\caption{\label{fig:sun-flow}a) angular velocity profile, $\Omega\left(r,\theta\right)/2\pi$,
contours are in range of 327-454 nHz; b) the radial profiles of the
angular velocity for latitudes: $\varphi=0^{\circ}$, $30^{\circ}$
and $60^{\circ}$; c) streamlines of the meridional circulation; d)
radial profile of the meridional flow at $\theta=45^{\circ}$.}
\end{figure}

Figure \ref{fig:sun-flow} shows profiles of the angular velocity,
streamlines of the meridional circulations and the radial profiles
of the angular velocity and the meridional flow velocity for a set
of latitudes. The given results were discussed in details by \citet{2018ApJ854.67P}.
The model shows the double-layer circulation pattern with the upper
stagnation point at $r=0.88R_{\odot}$. The amplitude of the surface
poleward flow is about 15m/s. The angular velocity profile shows a
strong subsurface shear that is higher at low latitudes and it is
less near poles. Contrary to results of \citet{Zhao13m} and model
of \citet{PK13} the double-cell meridional circulation structure
extends from equator to pole. This is partly confirmed by the new
results of helioseismology by \citet{2017ApJ849.144C} who also found
that the poleward flow at the surface goes close enough to pole. {It
is important no mention that the current results of the helioseismic
inversions for the meridional circulation remains controversial For
example, \citet{2015ApJ813.114R} found that the meridional circulation
can be approximated by a single-cell structure with the return flow
deeper than 0.77R$_{\odot}$. However, their results indicate an additional
weak cell in the equatorial region, and contradict to the recent results
of \citet{2017ApJ845.2B} who confirmed a shallow return flow at 0.9R$_{\odot}$.
Also, their results indicated that the upper meridional circulation
cell extends close to the solar pole.}

\subsection{Dynamo equations}

We model evolution of the large-scale axisymmetric magnetic field,
$\overline{\mathbf{B}}$, by the mean-field induction equation \citep{KR80},
\begin{equation}
\partial_{t}\overline{\mathbf{B}}=\boldsymbol{\nabla}\times\left(\boldsymbol{\mathcal{E}}+\mathbf{\overline{U}}\times\overline{\mathbf{B}}\right),\label{eq:mfe-1}
\end{equation}
where, $\boldsymbol{\mathcal{E}}=\left\langle \mathbf{u\times b}\right\rangle $
is the mean electromotive force with $\mathbf{u}$ and $\mathbf{b}$
standing for the turbulent fluctuating velocity and magnetic field
respectively.

Similar to our recent paper (see, \citep{2014ApJ_pipk,2017MNRAS.466.3007P}),
we employ the mean electromotive force in form: 
\begin{equation}
\mathcal{E}_{i}=\left(\alpha_{ij}+\gamma_{ij}\right)\overline{B}_{j}-\eta_{ijk}\nabla_{j}\overline{B}_{k}.\label{eq:EMF-1-1}
\end{equation}
where symmetric tensor $\alpha_{ij}$ models the generation of magnetic
field by the $\alpha$- effect; antisymmetric tensor$\gamma_{ij}$
controls the mean drift of the large-scale magnetic fields in turbulent
medium, including the magnetic buoyancy; the tensor $\eta_{ijk}$
governs the turbulent diffusion. The reader can find further details
about the $\boldsymbol{\mathcal{E}}$ in the above cited papers.

The $\alpha$ effect takes into account the kinetic and magnetic helicities
in the following form: 
\begin{eqnarray}
\alpha_{ij} & = & C_{\alpha}\eta_{T}\psi_{\alpha}(\beta)\alpha_{ij}^{(H)}+\alpha_{ij}^{(M)}\frac{\overline{\chi}\tau_{c}}{4\pi\overline{\rho}\ell^{2}},\label{alp2d-2}\\
\eta_{T} & = & \frac{\nu_{T}}{\mathrm{Pm_{T}}}
\end{eqnarray}
where $C_{\alpha}$ is a free parameter which controls the strength
of the $\alpha$- effect due to turbulent kinetic helicity; tensors
$\alpha_{ij}^{(H)}$ and $\alpha_{ij}^{(M)}$ express the kinetic
and magnetic helicity parts of the $\alpha$-effect, respectively;
$\mathrm{Pm_{T}}$ is the turbulent magnetic Prandtl number, and $\overline{\chi}=\left\langle \mathbf{a}\cdot\mathbf{b}\right\rangle $
($\mathbf{a}$ and $\mathbf{b}$ are the fluctuating parts of magnetic
field vector-potential and magnetic field vector). Both the $\alpha_{ij}^{(H)}$
and the $\alpha_{ij}^{(M)}$ depend on the Coriolis number. Function
$\psi_{\alpha}(\beta)$ controls the so-called ``algebraic'' quenching
of the $\alpha$- effect where $\beta=\left|\overline{\mathbf{B}}\right|/\sqrt{4\pi\overline{\rho}u'^{2}}$,
$u'$ is the RMS of the convective velocity. It is found that $\psi_{\alpha}(\beta)\sim\beta^{-3}$
for $\beta\gg1$. The $\alpha$- effect tensors $\alpha_{ij}^{(H)}$
and $\alpha_{ij}^{(M)}$ are given in Appendix. 
\begin{figure}
\includegraphics[width=0.9\columnwidth]{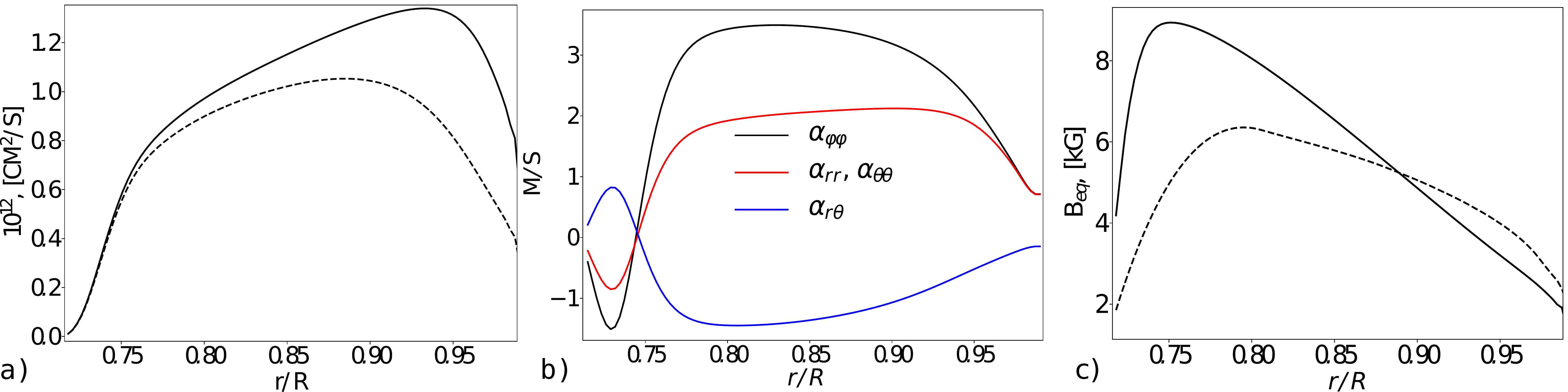}

\caption{\label{fig:alp} a) the radial profiles of the total (solid line)
and anisotropic (dashed line) parts of the eddy diffusivity at $\theta=45^{\circ}$;
b) radial profiles of the kinetic $\alpha$-effect components at $\theta=45^{\circ}$;
c) the equipartition strength of the magnetic field, $B_{eq}=\sqrt{4\pi\overline{\rho}u'^{2}}$,
where $u'$ is determined by the equatorial profile of the mean entropy,
see the Eq(\ref{eq:uc}); the dashed line is from results of the reference
model (MESA code) and the solid line is for the rotating convection
zone,i.e., after solution of the Eq(\ref{eq:heat}).}
\end{figure}

Contribution of the magnetic helicity to the $\alpha$-effect is expressed
by the second term in Eq.(\ref{alp2d-2}). The evolution of the turbulent
magnetic helicity density, $\overline{\chi}=\left\langle \mathbf{a}\cdot\mathbf{b}\right\rangle $,
is governed by the conservation law \citep{pip13M}:

\begin{eqnarray}
\frac{\partial\overline{\chi}}{\partial t} & = & -2\left(\boldsymbol{\mathcal{E}}\cdot\overline{\bm{B}}\right)-\frac{\overline{\chi}}{R_{m}\tau_{c}}+\boldsymbol{\nabla}\cdot\left(\eta_{\chi}\boldsymbol{\nabla}\bar{\chi}\right)\label{eq:hel-1}\\
 &  & -\eta\overline{\mathbf{B}}\cdot\mathbf{\overline{J}}-\boldsymbol{\nabla}\cdot\left(\boldsymbol{\mathcal{E}}\times\overline{\mathbf{A}}\right),\nonumber 
\end{eqnarray}
where $R_{m}=10^{6}$ is the magnetic Reynolds number and $\eta$
is the microscopic magnetic diffusion. In the drastic difference to
anzatz of \citet{kleruz82}, the Eq(\ref{eq:hel-1}) contains the
term $\left(\boldsymbol{\mathcal{E}}\times\overline{\mathbf{A}}\right)$.
It consists of the magnetic helicity density fluxes which result from
the large-scale magnetic dynamo wave evolution. The given contribution
alleviates the catastrophic quenching problem \citep{hub-br12,pip13M}.
Also the catastrophic quenching of the $\alpha$-effect can be alleviated
with help of the diffusive flux of the turbulent magnetic helicity,
$\boldsymbol{\boldsymbol{\mathcal{F}}}^{\chi}=-\eta_{\chi}\boldsymbol{\nabla}\bar{\chi}$
\citep{guero10,chatt11}. The coefficient of the turbulent helicity
diffusivity, $\eta_{\chi}$, is a parameter in our study. It affects
the hemispheric helicity transfer \citep{mitra10}.

In the model we take into account the mean drift of large-scale field
due to the magnetic buoyancy, $\gamma_{ij}^{(buo)}$ and the gradient
of the mean density, $\gamma_{ij}^{(\Lambda\rho)}$: 
\begin{eqnarray}
\gamma_{ij} & = & \gamma_{ij}^{(\Lambda\rho)}+\gamma_{ij}^{(buo)},\nonumber \\
\gamma_{ij}^{(\Lambda\rho)} & = & 3C_{pum}\eta_{T}\left(f_{1}^{(a)}\left(\mathbf{\boldsymbol{\Omega}}\cdot\boldsymbol{\Lambda}^{(\rho)}\right)\frac{\Omega_{n}}{\Omega^{2}}\varepsilon_{inj}-\frac{\Omega_{j}}{\Omega^{2}}\varepsilon_{inm}\Omega_{n}\Lambda_{m}^{(\rho)}\right)\label{eq:pump1}\\
\gamma_{ij}^{(buo)} & = & -\frac{\alpha_{MLT}u'}{\gamma}\beta^{2}K\left(\beta\right)g_{n}\varepsilon_{inj},\nonumber 
\end{eqnarray}
where $\mathbf{\boldsymbol{\Lambda}}^{(\rho)}=\boldsymbol{\nabla}\log\overline{\rho}$
; functions $f_{1}^{(a)}$ and $K\left(\beta\right)$ are given in
\cite{kp93,2017MNRAS.466.3007P}. The standard choice of the pumping
parameter is $C_{pum}=1$. In this case the pumping velocity is scaled
in the same way as the magnetic eddy diffusivity. In the presence
of the multi-cell meridional circulation, the direction and magnitude
of the turbulent pumping become critically important for the modelled
evolution of the magnetic field. It is confirmed in the direct numerical
simulations, as well (see, \cite{2018AA609A..51W}). For the standard
choice, the turbulent pumping is about an order of magnitude less than
the meridional circulation. For this case, explanation of the latitudinal
drift of the toroidal magnetic field near the surface faces a problem
(cf., \cite{PK13}). To study the effect of turbulent pumping we
introduce this parameter $C_{pum}$.

For the bottom boundary we apply the perfect conductor boundary conditions:
$\mathcal{E}_{\theta}=0,\,A=0$. The boundary conditions at the top
are defined as follows. Firstly, following ideas of \citet{1992AA256371M}
and \citet{pk11apjl} we formulate the boundary condition in the form
that allows penetration of the toroidal magnetic field to the surface:
\begin{eqnarray}
\delta\frac{\eta_{T}}{r_{e}}B+\left(1-\delta\right)\mathcal{E}_{\theta} & = & 0,\label{eq:tor-vac}
\end{eqnarray}
where $r_{e}=0.99R_{\odot}$, and parameter $\delta=0.99$. The magnetic
field potential in the outside domain is 
\begin{equation}
A^{(vac)}\left(r,\mu\right)=\sum a_{n}\left(\frac{r_{e}}{r}\right)^{n}\sqrt{1-\mu^{2}}P_{n}^{1}\left(\mu\right).\label{eq:vac-dec}
\end{equation}

The coupled angular momentum and dynamo equations are solved using
finite differences for integration along the radius and the pseudospectral
nodes for integration in latitude. The number of mesh points in radial
direction was varied from $100$ to $150$. The nodes in latitude
are zeros of the Legendre polynomial of degree ${N}$, where N was
varied from ${N=64}$ to ${N=84}$. The resolution with 64 nodes in
latitude and with 100 points in radius was found satisfactory. The
model employed the Crank-Nicolson scheme, using a half of the time-step
for integration in the radial direction and another half for integration
along latitude.

To quantify the mirror symmetry type of the toroidal magnetic field
distribution relative to equator we introduce the parity index $P$:
\begin{eqnarray}
P & = & \frac{E_{q}-E_{d}}{E_{q}+E_{d}},\label{eq:parity}\\
E_{d} & = & \int\left(B\left(r_{0},\theta\right)-B\left(r_{0},\pi-\theta\right)\right)^{2}\sin\theta d\theta,\nonumber \\
E_{q} & = & \int\left(B\left(r_{0},\theta\right)+B\left(r_{0},\pi-\theta\right)\right)^{2}\sin\theta d\theta,\nonumber 
\end{eqnarray}
where $E_{d}$ and $E_{q}$ are the energies of the dipole-like and
quadruple-like modes of the toroidal magnetic field at $r_{0}=0.9R_{\odot}$.
Another integral parameter is the mean density of the toroidal magnetic
field in the subsurface shear layer: 
\begin{equation}
\overline{B^{T}}=\sqrt{E_{d}+E_{q}}.\label{eq:bt}
\end{equation}
Another parameter characterize the mean strength of the dynamo processes
in the convection zone: 
\begin{equation}
\overline{\beta}=\left\langle \left|\overline{\mathbf{B}}\right|/\sqrt{4\pi\overline{\rho}u'^{2}}\right\rangle ,\label{eq:bet}
\end{equation}
where the averaging is done over the convection zone volume. The boundary
conditions Eq(\ref{eq:tor-vac}) provide the Poynting flux of the
magnetic energy out of the convection zone. Taking into account the
Eq(\ref{eq:flx}) the variation of the thermal flux at the surface
are given as follows: 
\begin{eqnarray}
\delta F & = & \delta F_{c}+\delta F_{B}\label{eq:dF}\\
\delta F_{c} & = & 4\frac{T_{e}}{T_{eff}}\frac{\delta\overline{s}}{c_{p}}\label{eq:flxb}\\
\delta F_{B} & = & \frac{1}{4\pi}\left(\mathcal{E}_{\phi}\overline{B}_{r}-\mathcal{E}_{\theta}\overline{B}_{\phi}\right),\label{eq:fcfb}
\end{eqnarray}
where $\delta\overline{s}$ is the entropy variation because of the
magnetic activity. The second term of the Eq(\ref{eq:dF}) governs
the magnetic energy input in the stellar corona.

\section{Results}

To match the solar cycle period we put $\mathrm{Pm_{T}}=10$ in all
our models. {The theoretical estimations of \citet{1994AN....315..157K}
gives $\mathrm{Pm_{T}}=4/3$. This is the long standing theoretical
problem of the solar dynamo period \citep{brsu05}. Currently, the
solar dynamo period can be reproduced for $\mathrm{Pm_{T}}\gg1$.
The issue exists both in the distributed and in the flux-transport
dynamo. Moreover, the flux-transport dynamo can reproduce the observation
only with the special radial profile of the eddy diffusivity (see,
e.g., \citep{2006ApJ...647..662R}). In our models we employ the rotational
quenching the eddy diffusivity coefficients and the high $\mathrm{Pm_{T}}$.
Figures \ref{fig:alp}a and b show the radial profiles of the eddy
diffusivity coefficients and components of the $\alpha_{ij}^{(H)}$
at latitude $45^{\circ}$ in our model for $\mathrm{Pm_{T}}=10$.
In the upper part of the convection zone the magnitude of the turbulent
magnetic diffusivity is close to estimations of \citet{1993AA...274..521M}
based on observations of the sunspot decay rate. The eddy diffusivity
is an order of $10^{10}$cm/s and less near the bottom of the convection
zone. The diffusivity profile is the same as in our previous paper
\citet{2014ApJ_pipk}. }

{The radial profiles of the $\alpha$ effect for $C_{\alpha}=C_{\alpha}^{(cr)}$
are illustrated in Figure \ref{fig:alp}b. The $\alpha$ effect (cf,
the above discussion about $\Lambda$-effect) change the sign near
the bottom of the convection zone. This is also found in the direct
numerical simulations \citep{2006AA455.401K}.}

Table \ref{tab:C} gives the list of our models, their control and
output parameters. We sort the models with respect to magnitude of
the $\alpha$ effect using the ratio ${\displaystyle \frac{C_{\alpha}}{C_{\alpha}^{(cr)}}}$,
the magnitude of the eddy-diffusivity of the magnetic helicity density,
${\displaystyle \frac{\eta_{\chi}}{\eta_{T}}}$, where $\eta_{T}=\nu_{T}/\mathrm{Pm_{T}}$,
and $\nu_{T}$ is determined from Eq(\ref{eq:nu}), and with respect
of the magnetic feedback on the differential rotation.{ The
$C_{pum}$ controls the pumping velocity magnitude (see, Eq\eqref{eq:pump1});
the parameter ${\displaystyle \frac{\Delta\Omega}{\Omega_{0}}}$ show
the relative difference of the surface angular velocity between the
solar equator and pole; the strength of the dynamo is characterized
by the range of the magnetic cycle variations of $\overline{\beta}$
(see, Eq\eqref{eq:bet}); the dynamo cycle period; the magnitude of
the surface meridional circulation. From the Table 1 we see that the
nonkinematic runs show the magnetic cycle variations of ${\displaystyle \frac{\Delta\Omega}{\Omega_{0}}}$
and the surface meridional circulation. }

Figure \ref{fig:alp}b shows the radial profiles of the equipartition
strength of the magnetic field, $B_{eq}=\sqrt{4\pi\overline{\rho}u'^{2}}$
in the solar convection zone for the reference model (non-rotating)
given by MESA code and in the rotating convective zone. In the rotating
convection zone, the mean-entropy gradient is larger than in the nonrotating
case. {This is because of the rotational quenching of the eddy-conductivity.
The magnitude of the convective heat flux is determined by the boundary
condition at the bottom of the convection zone and it remains the
same for the rotating (our model) and nonrotating (MESA code) cases.
Assuming that the convective turnover time is not subjected to the
rotational quenching, the reduction of the eddy conductivity because
of the rotational quenching is compensated by the increase of the
mean-entropy gradient.} This results in the increase of the parameter
$B_{eq}$.

{The increase of the RMS convective velocity in case of the
rotating convection zone seems to contradict the results of direct numerical
simulations of \cite{2016AA596A.115W}. This is
likely because of inconsistent assumptions behind the MLT expression
for the RMS convective velocity, see Eq(\ref{eq:uc}). The given issue
can affect the amplitude of the dynamo generated magnetic field near the
bottom of the convection zone. Our models operate in regimes where
$\left|B\right|\le B_{eq}$, and the substantial part of the dynamo
quenching is due to magnetic helicity conservation. Therefore the
given issue does not much affect our results.}

\begin{table}
\caption{\label{tab:C}Control and output parameters of the dynamo models.}
\begin{tabular}{>{\centering}p{1cm}>{\centering}p{2cm}>{\centering}p{1cm}>{\centering}p{2cm}>{\centering}p{1.5cm}>{\centering}p{2cm}>{\centering}p{1cm}>{\centering}p{1cm}}
\toprule 
Model  & $C_{pum}$  & ${\displaystyle \frac{\eta_{\chi}}{\eta_{T}}}$  & ${\displaystyle \frac{C_{\alpha}}{C_{\alpha}^{(cr)}}}$  & ${\displaystyle \frac{\Delta\Omega}{\Omega_{0}}}$  & $\overline{\beta}$  & \begin{centering}
Period 
\par\end{centering}
{[}YR{]}  & $\max U_{\theta}$

{[}M/S{]}\tabularnewline
\midrule 
M1  & 1  & $0.1$  & 1.1  & 0.279  & 0.05-0.1 & 6  & 15\tabularnewline
\midrule 
M2  & Pm$_{T}$  & 0.1  & 1.1  & 0.279  & 0.13-0.24  & 10.5  & 15\tabularnewline
\midrule 
M4  & -/-  & 0.3  & -/-  & 0.279  & 0.15-0.3  & 10.5,12.05  & 15\tabularnewline
\midrule 
M5  & -/-  & 0.01  & -/-  & 0.279 & 0.14-0.26  & 9.3  & 15\tabularnewline
\midrule 
 &  &  & Nonkinematic & runs &  &  & \tabularnewline
\midrule 
M3  & Pm$_{T}$  & 0.1 & 1.1  & 0.263-0.275  & 0.11-0.21  & 10.3  & 15.0

$\pm0.5$\tabularnewline
\midrule 
M3a2  & -/-  & -/-  & 2  & 0.232-0.253  & 0.36-0.66  & 4.7,265  & 14.0

$\pm2.1$\tabularnewline
\midrule 
M3a3  & -/-  & -/-  & 3  & 0.22-0.24  & 0.52-0.91  & 4.0  & 14.5 $\pm2.5$\tabularnewline
\midrule 
M3a4  & -/-  & -/-  & 4  & 0.21-0.245  & 0.68-1.05  & 3.4  & 14.7

$\pm3.$5\tabularnewline
\bottomrule
\end{tabular}
\end{table}

\subsection{Effects of turbulent pumping}

\begin{figure}
\includegraphics[width=0.8\columnwidth]{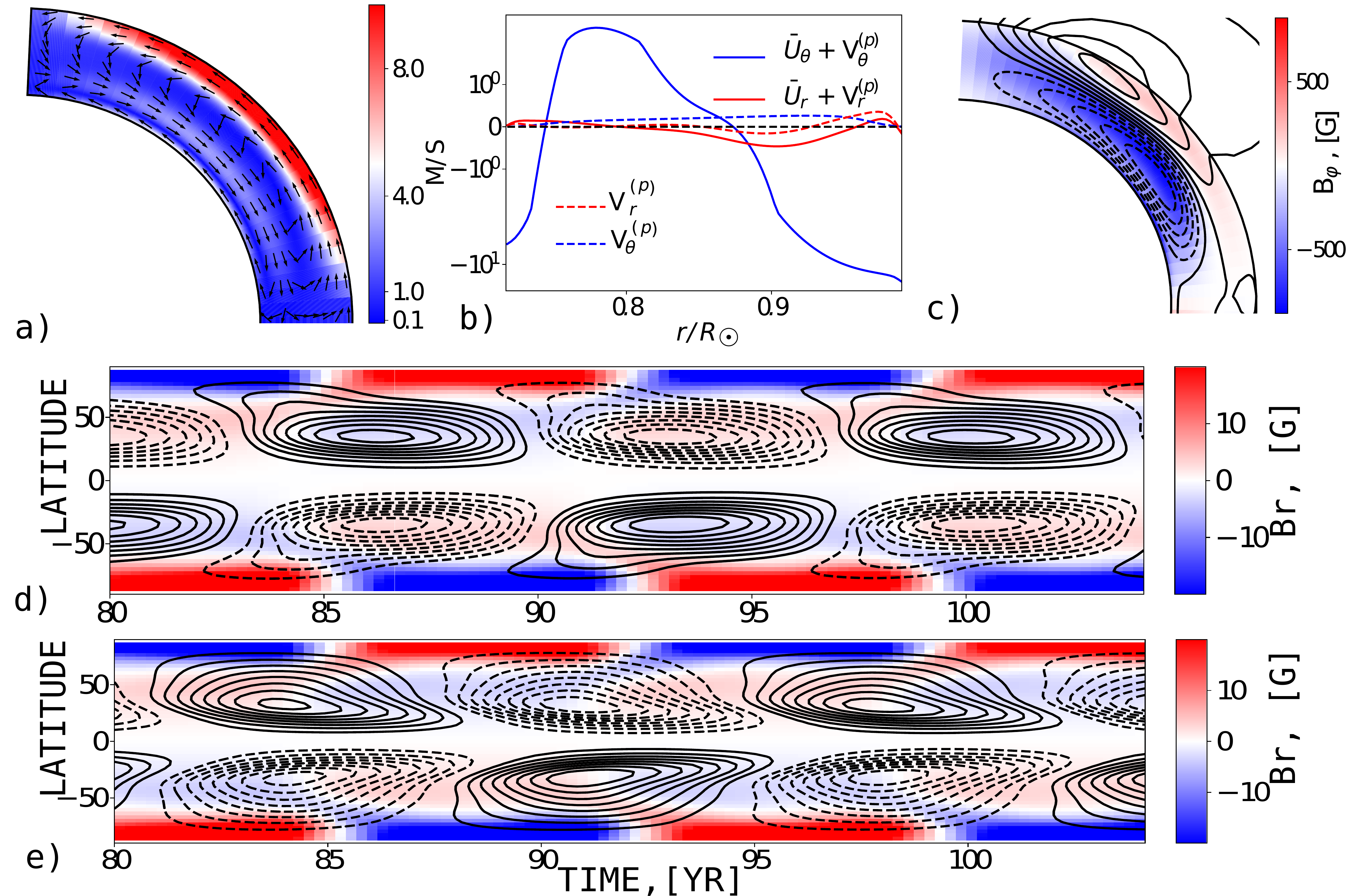}

\caption{\label{pumpa}a) Direction of pumping velocity of the toroidal magnetic
field in model M1; b) the effective velocity drift of the toroidal
magnetic field (pumping + meridional circulation); c) the snapshot
of the toroidal magnetic field distribution (color image) and streamlines
of the poloidal magnetic field in the Northern hemisphere of the Sun;
d) the time-latitude diagram of the toroidal magnetic field evolution
(contours in range of of $\pm500$G at $r=0.9R$ and radial magnetic
field at the surface (color image).}
\end{figure}

As the first step, we consider the kinematic dynamo model with the
nonlinear $\alpha$ effect. Results of the model M1 are shown in Figure
\ref{pumpa}. The model M1 roughly agree with results of \citet{PK13}
(hereafter, PK13). It employs the same mean electromotive force as
in our previous paper. In particular, the maximum pumping velocity
is the order of 1m/s. The effective velocity drift due to the magnetic
pumping and meridional circulation is shown in Figures \ref{pumpa}(a)
and (b). Figures \ref{pumpa}(d) and (e) show the time-latitude variations
of the toroidal magnetic field at $r=0.9R$ and in the middle of the
convection zone. The agreement with the solar observations is worse
than in the previous model PK13 because of difference in the meridional
circulation structure. The model of PK13 employed the meridional circulation
profile provided by results of \citet{Zhao13m}. In that profile,
the near-surface meridional circulation cell is more shallow and it
does not touch the pole as it happens in the present model, see, Figure
\ref{fig:sun-flow}. By this reason, the polar magnetic field in model
M1 is much larger than in results of PK13.

\begin{figure}
\includegraphics[width=0.8\columnwidth]{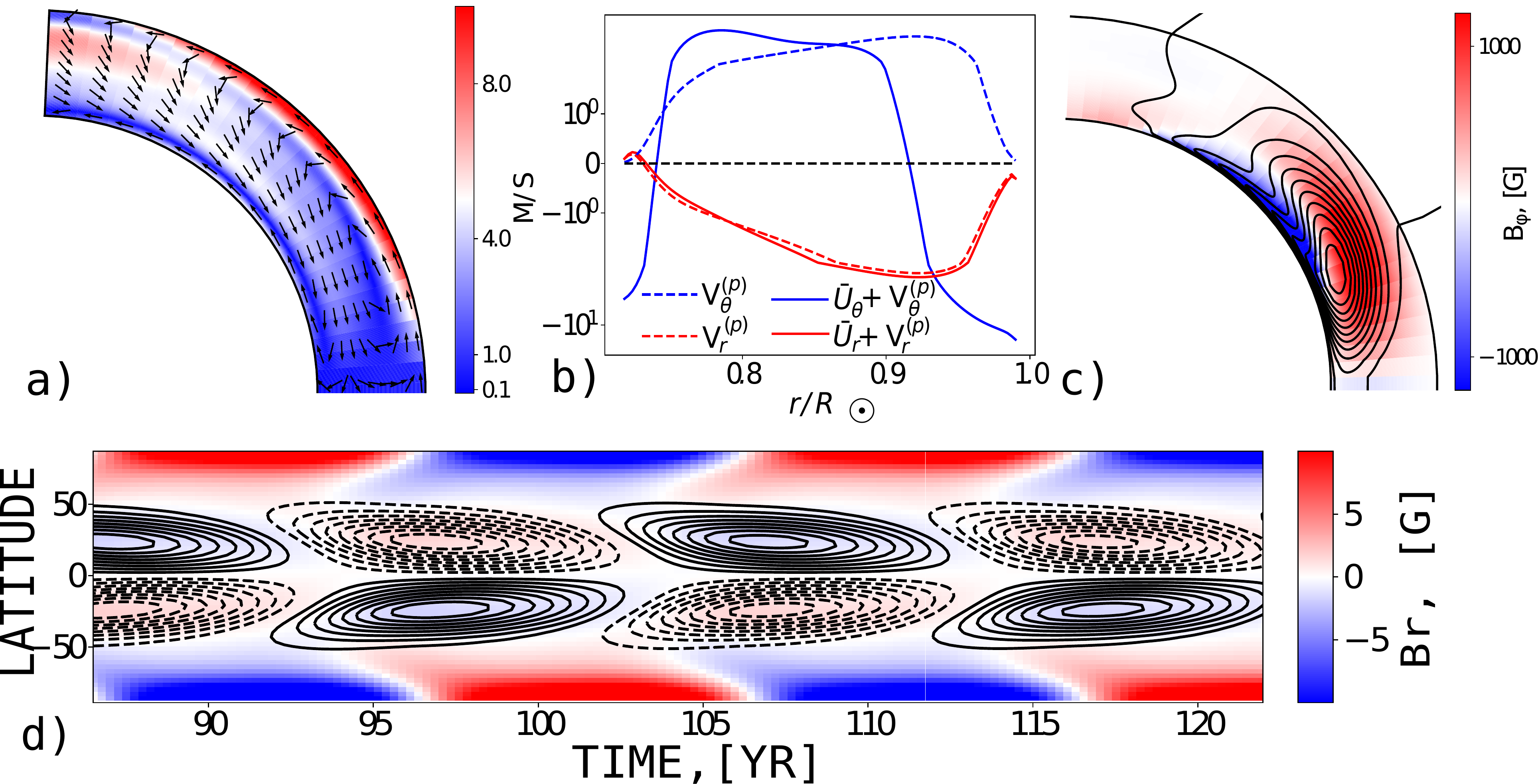}

\caption{\label{pumpb}a) Direction of pumping velocity of the toroidal magnetic
field in model M2; b) the effective velocity drift of the toroidal
magnetic field (pumping + meridional circulation); c) the snapshot
of the toroidal magnetic field distribution (color image) and streamlines
of the poloidal magnetic field in the Northern hemisphere of the Sun;
d) the time-latitude diagram of the toroidal magnetic field evolution
(contours in range of of $\pm1$kG at $r=0.9R$ and radial magnetic
field at the surface (color image).}
\end{figure}

For the purpose of our study, it is important to get the properties
of the dynamo solution as close as possible to results of solar observations.
To solve the above issues we increase the turbulent pumping velocity
magnitude by factor $\mathrm{Pm_{T}}$. The results are shown in Figure
\ref{pumpb}. The model has the correct time-latitude diagram of the
toroidal magnetic field in the subsurface shear layer. The surface
radial magnetic field evolves in agreement with results of observations
\citep{2013AARv2166S}. The magnitude of the polar magnetic field
is 10 G, which is in a better agreement with observations (e.g., \cite{2007AAS...210.2405L})
than the model M1. Figures \ref{pumpb} (a) and (b) show the effective
velocity drift of the large-scale toroidal magnetic field. The equatorward
drift with magnitude the  order of 1-2 m/s operates in major part of the
solar convection zone from $0.75R$ to $0.91R$. Interesting that
the obtained results are similar to those from the direct numerical
simulation of \citet{2018AA609A..51W}. Note that in the given model
the magnitude of the pumping velocity is about factor 2 less than
in results of \citet{2018AA609A..51W}. It seems that some of the
issues in model M1 would be less pronounced if the meridional circulation
pattern was closer to results of helioseismology of \citet{Zhao13m}
or \citet{2017ApJ849.144C}. However results of the direct numerical
simulations of \citep{2016AA596A.115W,2018AA609A..51W} seem to show
that the evolution of the large-scale magnetic field inside convection
zone does not depend much on the meridional circulation. This argues
for the strong magnetic pumping effects in the global dynamo. This
important issue can be debated further. This is out of the main scopes
of this paper. The rest of our models employ the same pumping effect
as in the model M2 (see, Table \ref{tab:C}). 

\subsection{The global flows oscillations in magnetic cycle}

\begin{figure}
\includegraphics[width=0.8\columnwidth]{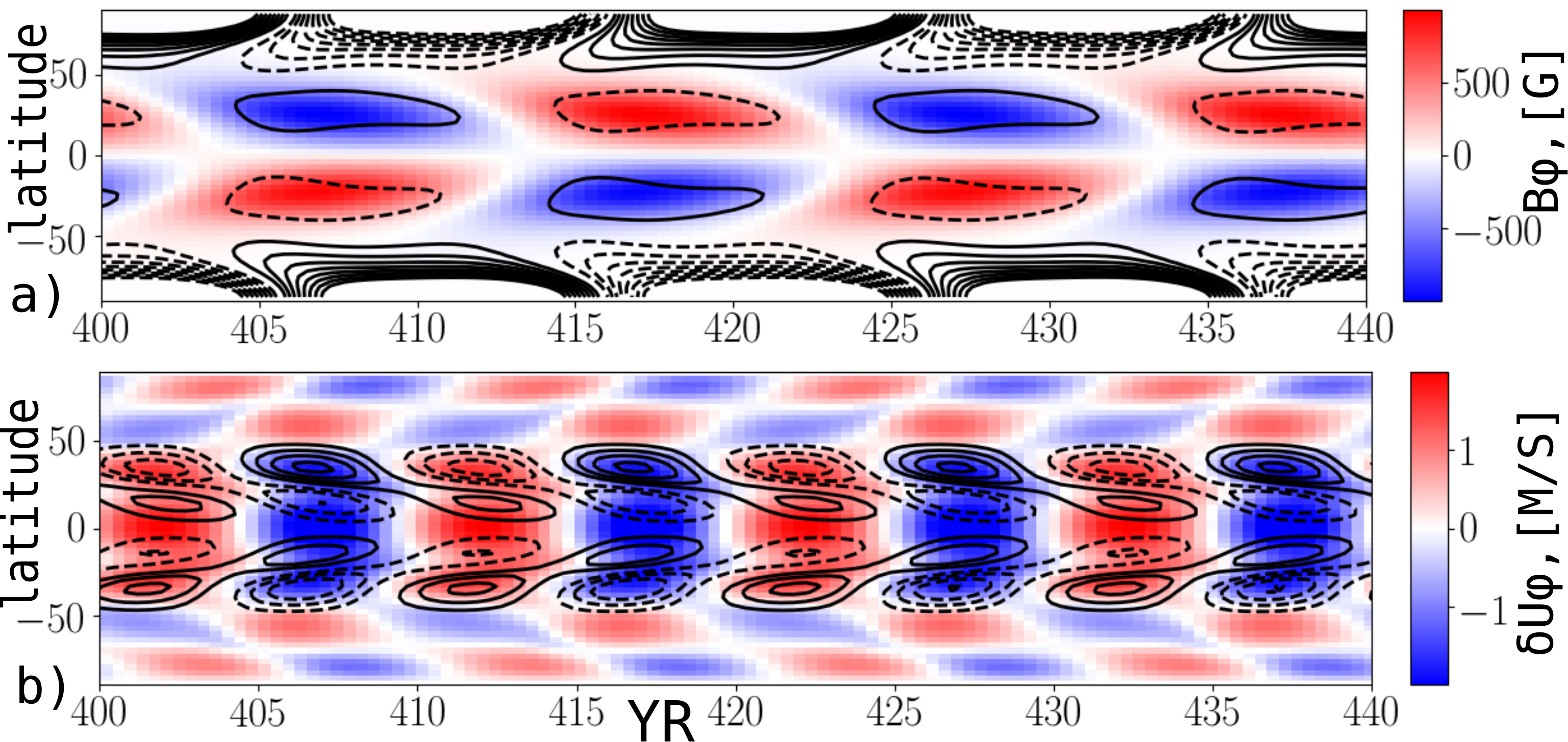}

\caption{\label{fig:s-bat}The model M3, a) Time-latitude butterfly diagram
for the toroidal field in the upper part of the convection zone (color
image) and the surface radial magnetic field shown by contours ($\pm$5G);
b) the surface variations of the azimuthal velocity (color image)
and the meridional velocity (contours in the range of $\pm0.5$ m/s). }
\end{figure}

Figure \ref{fig:s-bat} show the time-latitude diagrams of the magnetic
field and the global flow variations for the model M3. The torsional
oscillations on the surface are about $\pm2$m/s. They are defined
as follows, $\delta U_{\phi}=\left(\Omega\left(r,\theta t\right)-\overline{\Omega\left(r,\theta,t\right)}\right)r\sin\theta$,
where the averaging is done over the stationary phase of evolution.
The torsional wave has both the equator- and poleward branches. In
the equatorward torsional wave, the change from the positive to negative
variation goes about 2 years ahead of the maxima of the toroidal magnetic
field wave. This agrees with results of observations of \citet{2011JPhCS271a2074H}
and with direct numerical simulations of \citet{2016ApJ828L.3G}.
The magnitude of the meridional flow variations agrees with results
of \citet{2014ApJ789L7Z}. Also, we see that on the surface the meridional
velocity variations converge toward the maximum of the toroidal magnetic
field wave. This is also in qualitative agreement with the observations.

\begin{figure}
\includegraphics[width=0.95\columnwidth]{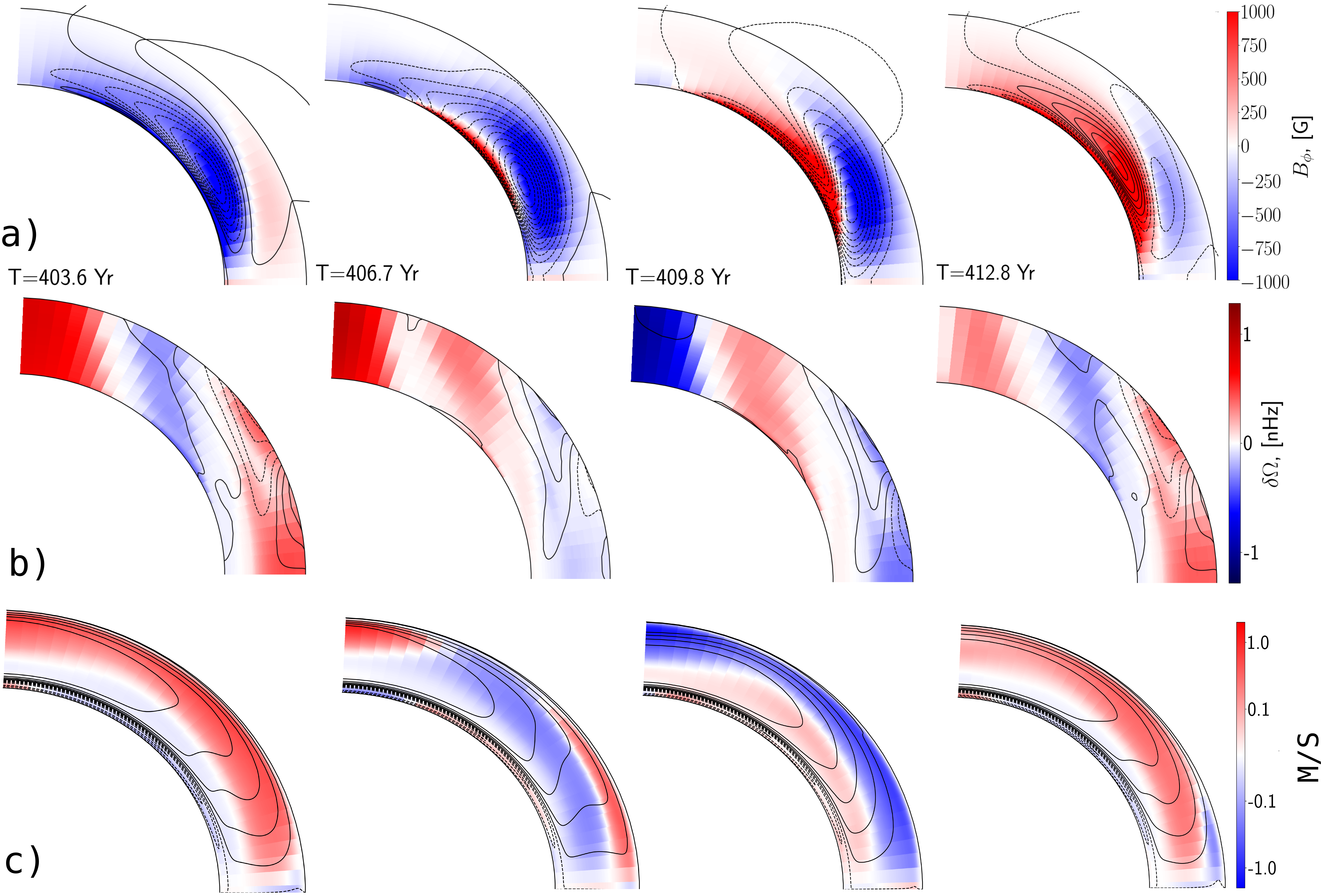}\caption{\label{fig:M3}The model M3: a) snapshots of the magnetic field in
four phase of the magnetic cycle, the toroidal magnetic field strength
is shown by color, contours show streamlines of the poloidal magnetic
field; b) color image show variations of the angular velocity, contours
(range of $\pm0.5$m/s) show variations of the meridional flow; c)
contours show the azimuthal component of the total (kinetic and magnetic
helicity parts)$\alpha$ -effect, the background image shows the part
of the $\alpha_{\phi\phi}$ induced by the magnetic helicity conservation
(see, the second term of Eq(\ref{alp2d-2})).}
\end{figure}

Figure \ref{fig:M3} shows snapshots of the magnetic field, the global
flows variations and the azimuthal component of the total (kinetic
and magnetic helicity parts)$\alpha$ -effect for a half magnetic
cycle. The Figure shows that a new cycle starts at the bottom of the
convection zone. The main part of the dynamo wave drifts to surface
equatorward. There is a polar branch which propagates poleward along
the bottom of the convection zone. The torsional oscillations, as
well as, the meridional flow variations are elongated along the axis
of rotation. This can be interpreted as a result of mechanical perturbation
of the Taylor-Proudman balance \citep{2006ApJ...647..662R}. We postpone
the detailed analysis of the torsional oscillation to another paper.
{Figure \ref{fig:M3}b shows that maxima of the meridional
flow variations are located at the upper boundary of the dynamo domain.}
This is because the main drivers of the meridional circulation, which
are the baroclinic forces, have the maximum near the boundaries of
the solar convection zone \citep{rem2005ApJ,2015ApJ...804...67F,2017arXiv170202421P}.
In comparing Figures \ref{fig:M3}c and \ref{fig:alp} it is seen
that the dynamo wave affect the $\alpha$ -effect. Also, in agreement
with our previous model \citep{pip2013ApJ}, we find that the magnetic
helicity conservation results into increasing the $\alpha$ -effect
in the subsurface shear layer. It occurs just ahead of the dynamo
wave drifting toward the top. The given effect support the equatorward
propagation of the large-scale toroidal field in subsurface shear
layer \citep{kap2012}.

\begin{figure}
\includegraphics[width=0.8\columnwidth]{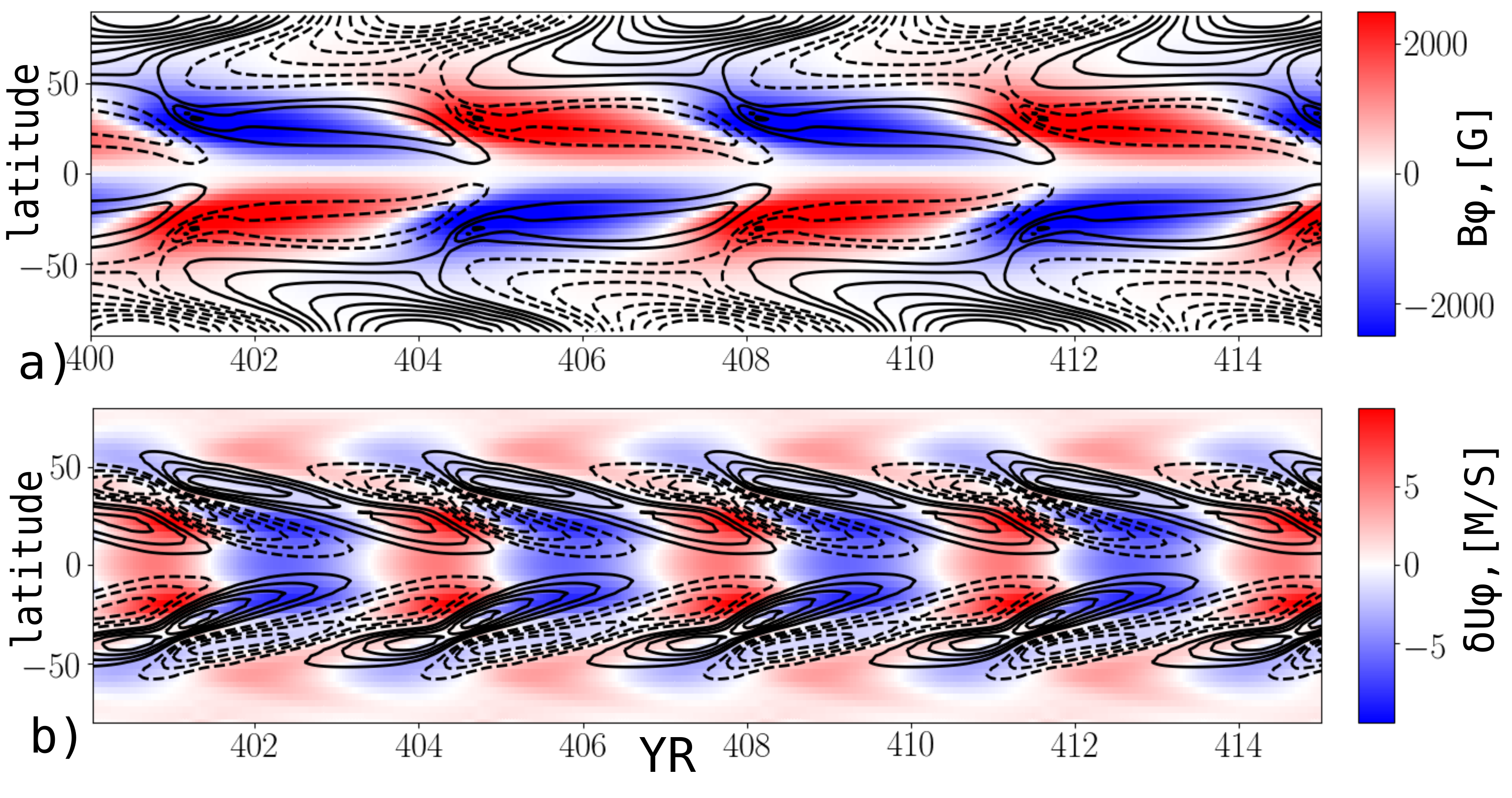}

\caption{\label{fig:m8}The model M3a4, a) Time-latitude butterfly diagram
for the toroidal field in the upper part of the convection zone (color
image) and the surface radial magnetic field shown by contours ($\pm$100G);
b) the surface variations of the azimuthal velocity (color image)
and the meridional velocity (contours in the range of $\pm3.5$ m/s). }
\end{figure}

The increasing the $\alpha$-effect parameter results in a number
of consequences for the non-linear evolution of the large-scale magnetic
field. The dynamo period is decreasing with the increase of the $\alpha$-effect
\citep{pk11}. The magnitude of the dynamo wave increases with the
increase of the parameter $C_{\alpha}$.{ Therefore, our models
show that in the distributed solar-type dynamo the dynamo period can
decrease with the increase of the magnetic activity level. This is
in agreement with the results of the stellar activity observations
of \citet{1984ApJ...287..769N,2009AA_strassm,2017PhDT3E}. Here we
for the first time demonstrate this effect in the distributed dynamo
model with the meridional circulation. }Figure \ref{fig:m8} shows
results for the model M3a4 with ${\displaystyle C_{\alpha}=4C_{\alpha}^{(cr)}}$.
The model shows the solar-like dynamo waves in the subsurface shear
layer. The toroidal magnetic field reaches the strength of 3kG in
the upper part of the convection zone. Simultaneously, the polar magnetic
field has the maximum strength of 100 G. Variations of the zonal and
meridional flows on the surface are about of factor 6 larger than
in the model M3. The model M3a4 show a high level magnetic activity
with a strong toroidal magnetic field in the subsurface layer and
very strong polar field. {Results of stellar observations show
that this is expected on the young solar analogs and late K-dwarfs
as well (e.g., \citep{2005LRSP2.8B,2016MNRAS1129S}). However, the
given results can not be consistent with those cases because in our
model the rotation rate is much slower than for the young solar-type
stars. Results of the linear models show that the internal differential
rotation and meridional circulation change with increase of the stellar
angular velocity \citep{kit11}.}

\begin{figure}
\includegraphics[width=0.7\columnwidth]{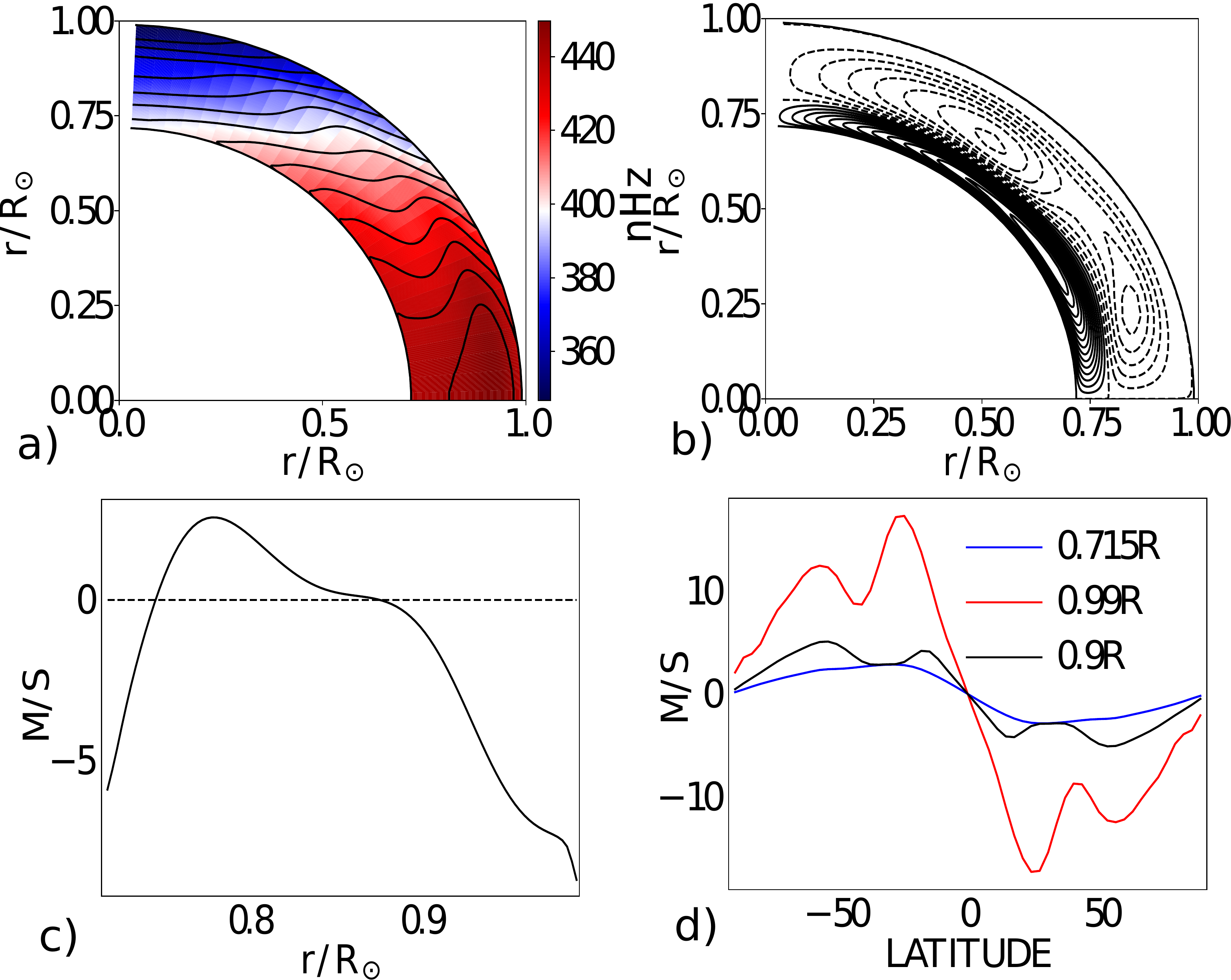}

\caption{\label{fig:M3a4dr}The model M3a4: a) the snapshot of angular velocity
profile, $\Omega\left(r,\theta\right)/2\pi$, contours are in range
of 347-450 nHz; b) the streamlines of the meridional circulation;
c) the radial profile of the meridional flow at $\theta=45^{\circ}$;
d) the profiles of the meridional flow at the specific depths of the
solar convection zone.}
\end{figure}

Figure \ref{fig:M3a4dr} shows snapshots of the global flows distributions
in the solar convection zone. In drastic difference to the model M3,
the counter-clockwise meridional circulation cell in the upper part
of the convection zone is divided into two parts. Also, the stagnation
point of the bottom cell is shifted equatorward.

\begin{figure}
\includegraphics[width=0.95\columnwidth]{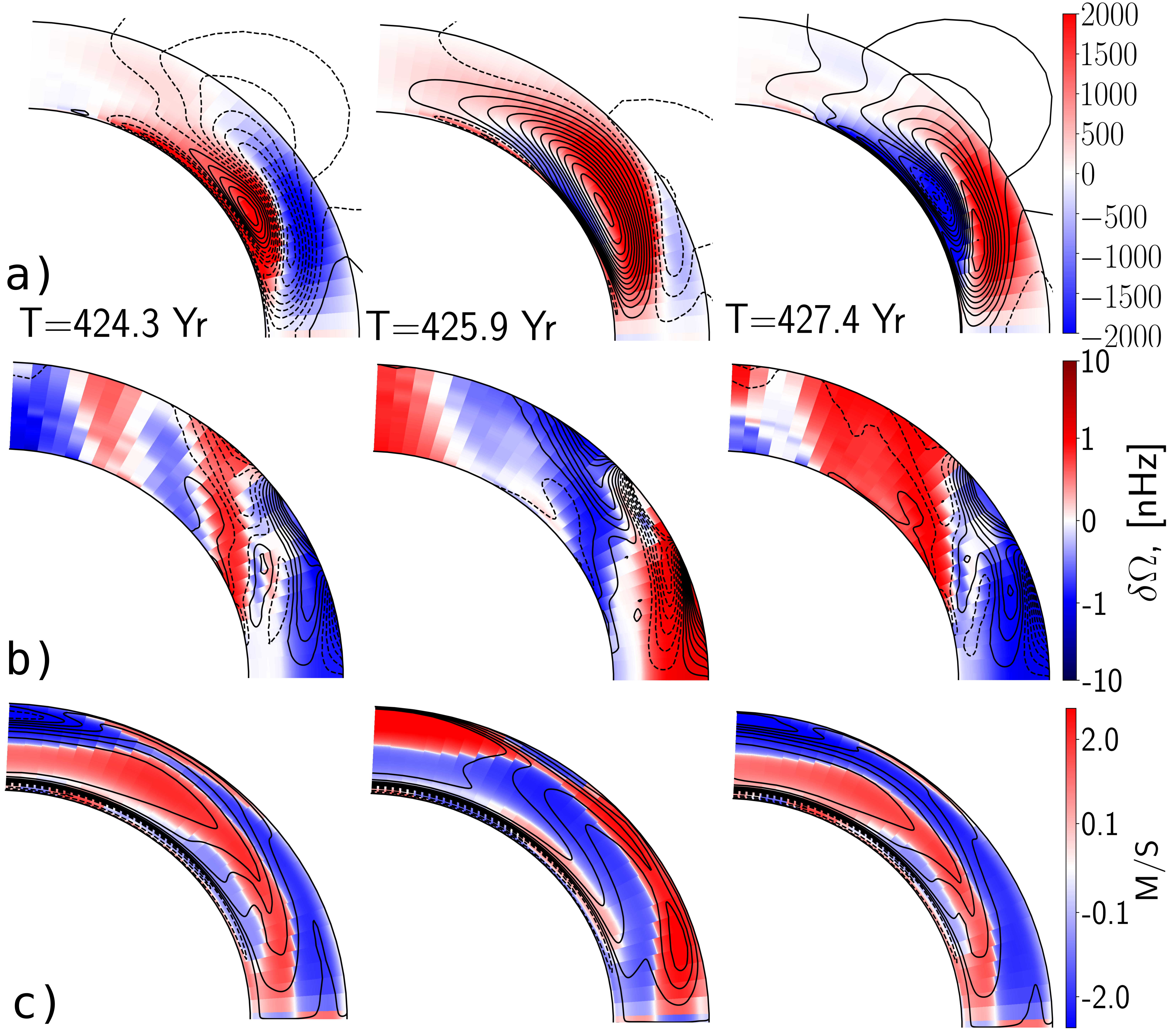}\caption{\label{fig:M3a4}The same as Figure \ref{fig:M3} for the model M3a4.}
\end{figure}

Figure \ref{fig:M3a4} shows the magnetic cycle variations of the
global flows in the Northern segment of the solar convection zone
for the model M3a4. The patterns of these variations are qualitatively
the same as in the model M3. The model M3a4 show the strong magnetic
cycle variations of the $\alpha$ effect. Similar to the model M3
we see that the magnetic helicity conservation results into increasing
the $\alpha$ -effect in the subsurface shear layer. It occurs just
ahead of the dynamo wave drifting toward the top. In the polar regions,
the $\alpha$ -effect inverses the sign during inversion of the polar
magnetic field. This is different from the model M3 which has the smaller
strength of the polar magnetic field than the model M3a4.

\begin{figure}
\includegraphics[width=0.8\columnwidth]{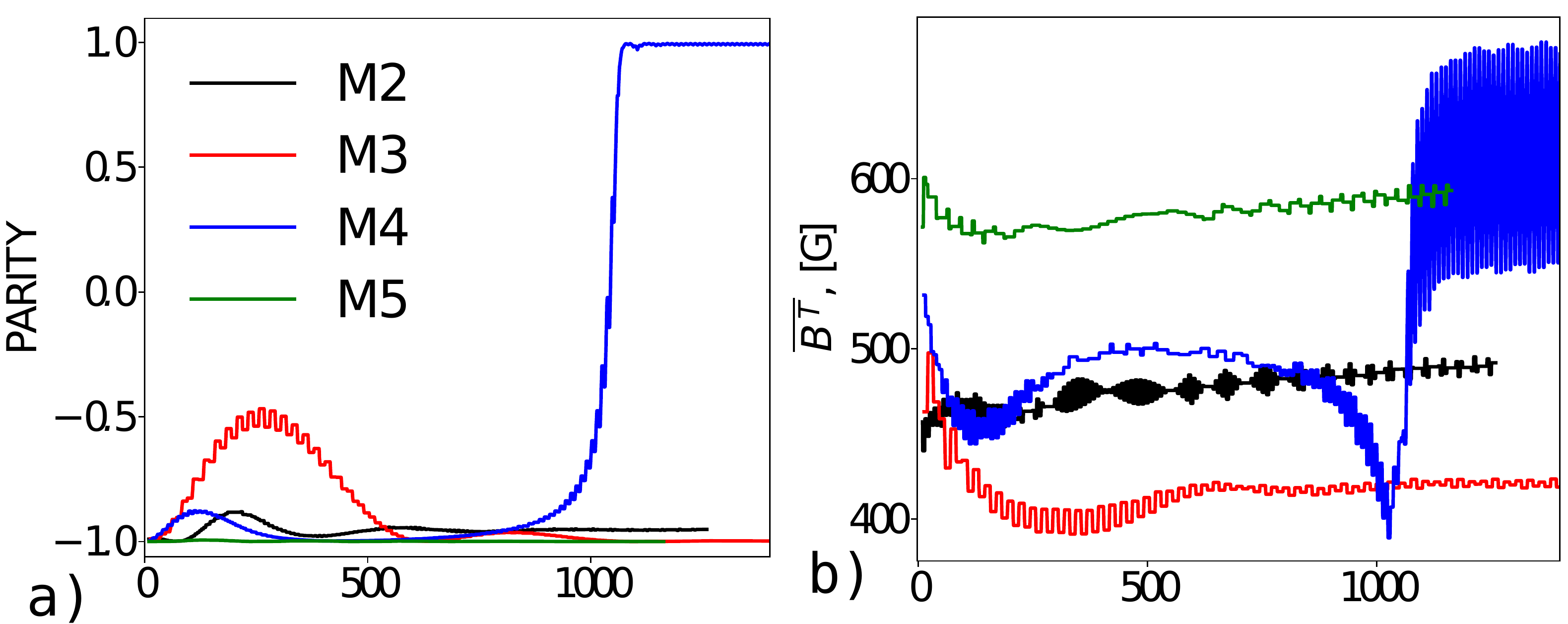}

\caption{\label{fig:avar}a) Variations of the equatorial symmetry (parity
index, see, Eq(\ref{eq:parity})); b) the same as (a) for the mean
density of the toroidal magnetic field flux in the subsurface shear
layer. The time series were smoothed to filter out the basic magnetic
cycle.}
\end{figure}

\subsection{The long-term dynamo evolution}

Figure \ref{fig:avar} shows the smoothed time series of evolution
of the global properties of the dynamo model, such as the equatorial
symmetry index, or the parity index $P$, (see, Eq(\ref{eq:parity}))
and the mean density of the toroidal magnetic field flux, $\overline{B^{T}}$,
in the subsurface shear layer, see, Eq(\ref{eq:bt}). In each time
series, the basic magnetic cycle was filtered out. The set of models
shown in Figure (\ref{fig:avar}) illustrates the effect of variations
of magnitude of the eddy-diffusivity of the magnetic helicity density,
${\displaystyle \frac{\eta_{\chi}}{\eta_{T}}}$. From results of \citet{mitra10},
it is expected that ${\displaystyle \frac{\eta_{\chi}}{\eta_{T}}}<1$.
In our set of models it is $0.01<{\displaystyle \frac{\eta_{\chi}}{\eta_{T}}}<0.3$.
The magnetic helicity diffusion affects the magnetic helicity exchange
between hemispheres \citep{2000JGR...10510481B,mitra10}. Therefore
it affects an interaction of the dynamo waves through the solar equator.
It is found that the increasing of ${\displaystyle \frac{\eta_{\chi}}{\eta_{T}}}$
results into change of the parity index $P$. The model M4 show the
symmetric about equator magnetic field. The magnitude of $\overline{B^{T}}$
in the model M4 is larger than in the models M3 and M5. Both models
M3 and M5 operate in a weak nonlinear regime with $0.13<\overline{\beta}<0.24$
(see, Eq(\ref{eq:bet}) and Table(\ref{tab:C})). The model M4 has
a slightly higher $\overline{\beta}$. This means that the magnetic
helicity diffusion affects the strength of the dynamo. This is in
agreement with \citet{guero10}. The smoothed time series of $\overline{B^{T}}$
in the model M4 show oscillations at the end of the evolution. This
is because the period of the symmetric dynamo mode ( $P=1$) is about
12 years that is larger than the period of the basic magnetic cycle
for antisymmetric mode $\left(P=-1\right)$ . The latter is about
10.5 years.

It is interesting to compare the kinematic and nonkinematic dynamo
models, which are the models M2 and M3. The nonkinematic model M3
has the smaller parameters $\overline{B^{T}}$and $\overline{\beta}$
than the model M2. Also, it is found that in the kinematic model M2
the stationary phase of evolution is the mix of the dipole-like and
quadrupole-like parity, with the mean $P\approx-0.9$. In the model
M3 the mean $P\approx-1$. Therefore the nonkinematic dynamo regimes
affect the equatorial symmetry of the dynamo solution \citep{bran89,p99}.
{Figure \ref{fig:avar}b shows another interesting difference
between the nonlinear kinematic and nonkinematic runs. The kinematic
models M2, M4 and M5 show a very slow evolution toward the stationary
stage. The given effect was reported earlier by \citet{pip2013ApJ}.
The high $R_{m}=10^{6}$ and small diffusivity $\eta_{\chi}$ result
to the long time-scale of establishment of the nonlinear balance in
magnetic helicity density distributions.}

\begin{figure}
\includegraphics[width=0.8\columnwidth]{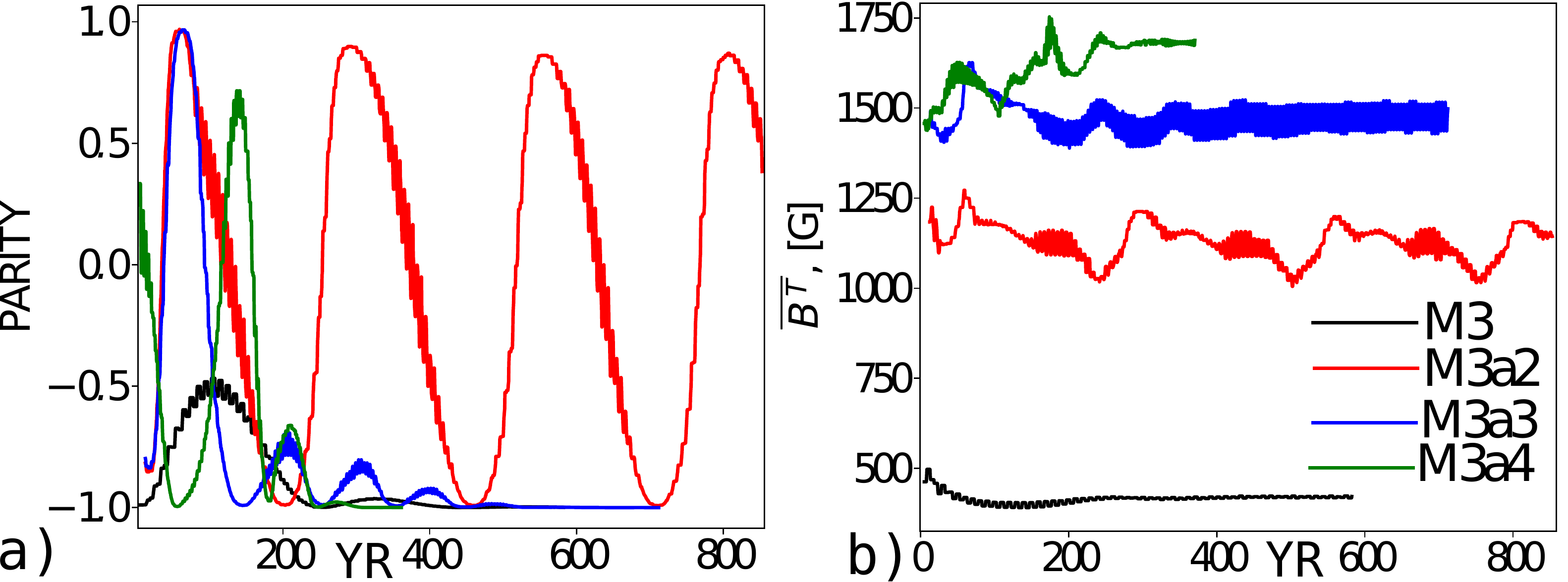}

\caption{\label{fig:lt}The same as Figure (\ref{fig:avar}) for the models
with different $C_{\alpha}$, see Table(\ref{tab:C}).}
\end{figure}

Figure (\ref{fig:lt}) shows results for the nonkinematic dynamo models
in a range the $\alpha$-effect parameter $C_{\alpha}$. The increasing
of the $C_{\alpha}$ results into increasing the nonlinearity of the
dynamo model. The parameter $\overline{\beta}$ grows from $0.2$
to $1$ with the increasing of $C_{\alpha}$by factor 4. The model
M3a2 shows the long-term periodic variations of the parity index and
the magnitude of the toroidal magnetic field $\overline{B^{T}}$.
These long-term cycles are likely due to the parity breaking because
of the hemispheric magnetic helicity exchange. We made the separate
run where the magnetic helicity conservation was ignored and did not
find the long-term cycles solution. These cycles are not robust against
changes of $C_{\alpha}.$ For the case $\eta_{\chi}=0.1\eta_{T}$,
they exist in the range $1.5C_{\alpha}^{(cr)}<C_{\alpha}<3C_{\alpha}^{(cr)}$.
The given range is likely to be changed with the change of the magnetic
helicity diffusion. We do not consider this case in our paper. Noticeably,
that the period of the long-cycle is likely changed with the variation
of the $\alpha$-effect. For the model M3a2, the period of the long
cycle is about 265 years. In the model M3a3, the establishment of
the stationary stage proceeds with the long-term oscillations of about
100 years period. We made the separate runs for the kinematic models
with the same $\alpha$-effect and magnetic helicity diffusivity parameters
as in the set shown in Figure (\ref{fig:lt}). It was confirmed that
the long-term cycles are exists in the range $1.5C_{\alpha}^{(cr)}<C_{\alpha}<3C_{\alpha}^{(cr)}$.
Therefore, the magnetic helicity evolution is the most important parameter
which governs the long-term periodicity in our model. The magnetic
parity breaking and its effect to the Grand activity cycle are lively
debated in the literature \citep{bran89,sok1994AA,1998MNRAS.297.1123K,2014JGRA119.6027F}.

\subsection{Magnetic cycle in the mean-field heat transport}

The global thermodynamic parameters are subjected to the magnetic
cycle modulation because of energy loss and gain for the magnetic
field generation and dissipation. Also, the magnetic field quenches
the magnitude of the convective heat flux. This affects the mean entropy
gradient. The Eq(\ref{eq:heat}) takes these processes into account.
We remind that in derivation of the mean-field heat transport equation
(see, \citep{2000ARep...44..771P}), it was assumed that the magnetic
and rotational perturbations of the reference thermodynamic state
are small. Also, the basic parameters of the reference state, like
$\overline{\rho}$, $\overline{T}$, etc, are given by the reference
stellar interior code (MESA). Another approach was suggested recently
by \citet{2015JPlPh..81e3904R}.

\begin{figure}
\includegraphics[width=0.5\columnwidth]{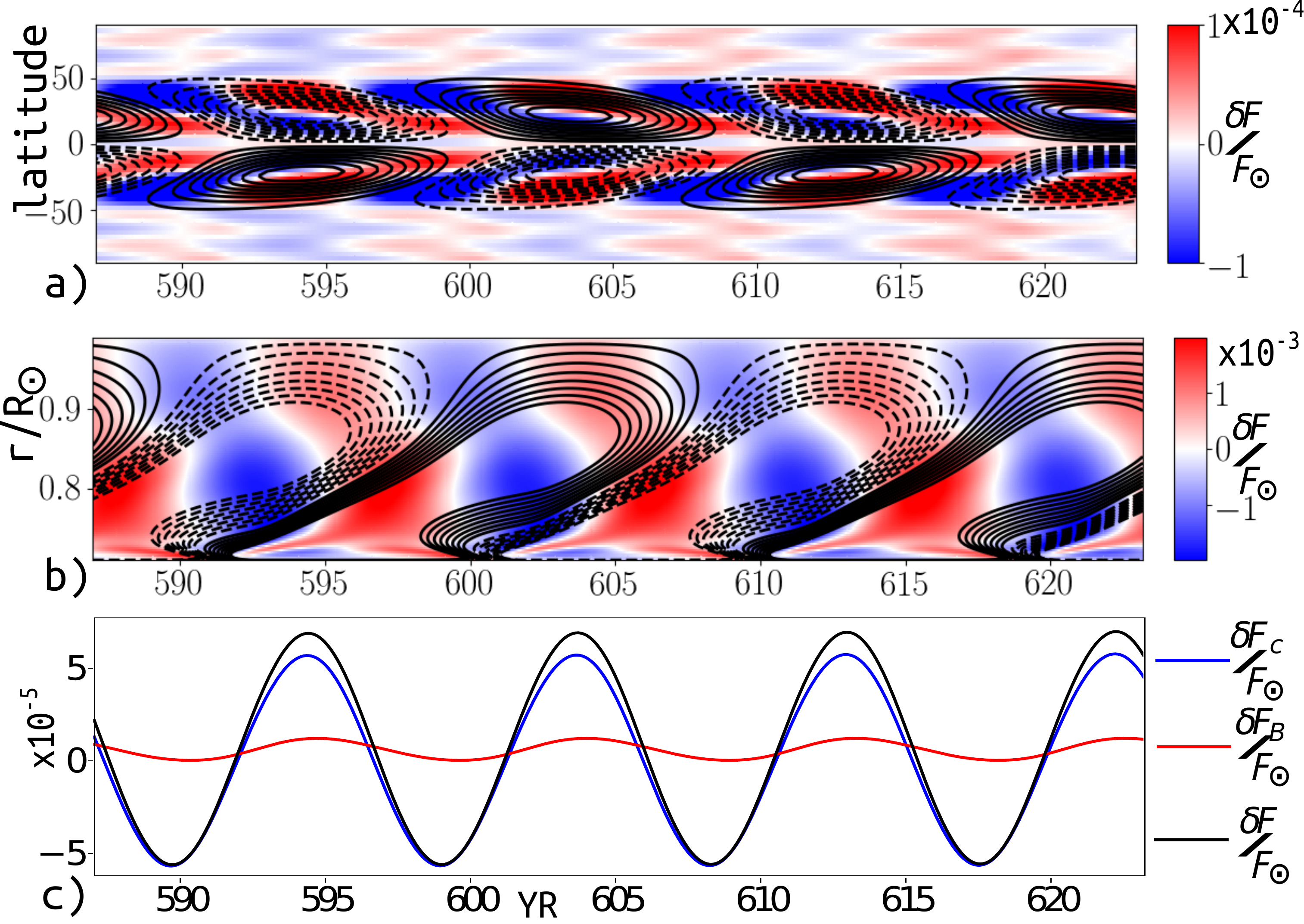}

\caption{\label{fig:m3h}Model M3, a) the time-latitude diagram for the near-surface
toroidal magnetic field (contours in range of $\pm1kG$) and variation
of the energy flux on the surface which is calculated relative to
the mean background convective flux; b) the same as (a) for the time-radius
diagram; c) the mean over the surface outflux of the convective energy
(blue line), the magnetic energy and the total energy (black line).}
\end{figure}

\begin{figure}
\includegraphics[width=0.5\columnwidth]{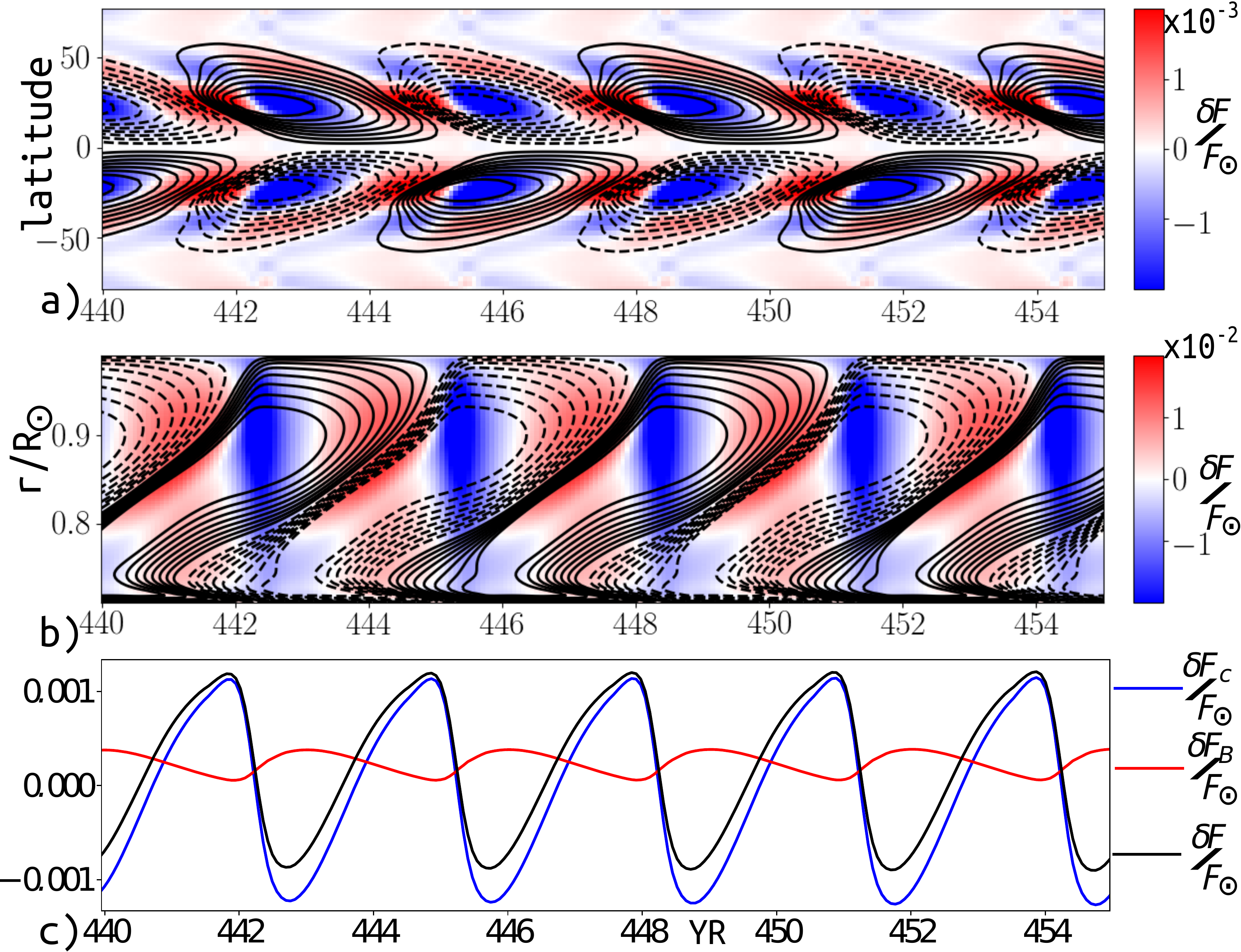}

\caption{\label{fig:m34}The same as Figure \ref{fig:m3h} for the model M3a4.
The contours of the toroidal magnetic field show the range $\pm3$kG. }
\end{figure}

The model calculates the mean entropy variations as an effect of the
global flows and magnetic activity. Some results in this direction
were previously considered in the dynamo models by \citet{1992AA...265..328B}
and \citet{2000ARep...44..771P}. Figures \ref{fig:m3h} and \ref{fig:m34}
show the time-latitude and time-radius diagrams for the magnetic field
evolution and variations of the convective heat flux, as well as the
mean heat and magnetic energy flux on the surface for the models M3
and M3a4. Evolution of the thermodynamic perturbations goes in form
of the cyclic modulations with the double frequency of the magnetic
cycle. Like in our previous studies we can identify two major sources
of the thermodynamic perturbations in the magnetic cycle. One is the
so-called magnetic shadow, which is due to the magnetic quenching
of the convective heat flux. This effect is quite small in comparison
to the total convective flux. Another effect is the energy expenses
on the large-scale dynamo. This contribution is governed by the term
${\displaystyle -\frac{1}{4\pi}\boldsymbol{\mathcal{E}}\cdot\nabla\times\boldsymbol{\overline{B}}}$in
the heat transport Eq(\ref{eq:heat}). Note that our boundary conditions
(see, Eq(\ref{eq:tor-vac})), beside the penetration of the toroidal
magnetic field to the surface, allow the energy flux from the dynamo
region. This flux is related to the strength of the toroidal magnetic
field on the surface.

The model M3 shows that the perturbation of the heat energy flux on
the surface goes in phase with the evolution of the near-surface toroidal
magnetic field. The decreasing of the heat flux corresponds to the
increase of the toroidal magnetic field. In the growing phase of the
cycle, the thermal energy is expended for the magnetic field generation.
The opposite process goes during the declining phase of the magnetic
cycle. The magnitude of the thermal flux perturbation is about $10^{-4}$
at the surface and it reaches $10^{-3}$ in the depth of the convection
zone. The cyclic perturbations of the mean energy flux on the surface
are about $10^{-4}$, which is an order of magnitude smaller than
in the solar observations.

The mean Poynting flux on the surface has a maximum order of $10^{-5}$
of the background heat energy flux. This flux represents the magnetic
energy input to the stellar corona. It can be considered as a part
of energy source for the magnetic cycle variation of the solar X-ray
luminosity. The solar observations show that variations the X-ray
background flux is the order of $10^{-6}$ \citep{2014ApJ793L.45W}.
{Therefore the magnetic energy flux in the weakly nonlinear
model M3 is enough to explain the solar X-ray luminosity variations
assuming that all magnetic energy input is transformed into soft X-ray
flux energy.}

The model M3a4 shows a different evolution of the thermal perturbations
in the solar cycle. The growing phase of the magnetic cycle is much
shorter than the declining phase. Also, the strength of the magnetic
field in the model M3a4 is about factor 3 higher than in the model
M3. The effect of the magnetic shadow dominates contributions from
the heat energy expenses to the dynamo. In this case, the minima of
the thermal energy flux are roughly located in the extremes of the
toroidal magnetic field. By this reason, the mean surface heat energy
flux reach minimum just after the maximum of the magnetic cycle. The
model M3a4 shows the relative variations of the mean energy flux of
the order $2\times 10^{-3}$, which is $0.2$ percent of the background
heat flux.{ The strong magnetic feedback of the heat flux in
the model M3a4 likely results to the origin of the second near-equatorial
meridional circulation cell.} We postpone the analysis of this effect
to another paper.

The model M3a4 shows the mean Poynting flux the order of $4\cdot10^{-4}$.
This value is comparable with the X-ray luminosity variations on the
young solar analogs, e.g., $\chi^{1}$Ori shows log$L_{x}\sim29$
\citep{1998ASPC154.1041G}, which is about $10^{-4}$ of its bolometric
luminosity.

\section{Discussion and conclusions}

The paper presents the new non-kinematic mean-field dynamo model which
takes into account the magnetic feedback on the angular momentum and
heat transport inside the convection zone. For the first time, the
dynamo model takes into account the complicated structure of the meridional
circulation cell in the convection zone. In following to \citep{2018ApJ854.67P},
the double-cell meridional circulation structure results from inversion
of the $\Lambda$-effect sign in the lower part of the convection
zone. The similar results are suggested by the helioseismology inversions.
Note that the recent results of \citet{2017ApJ849.144C} showed the
much smaller magnitude of the meridional circulation at the bottom
in comparison to the previous results of \citet{Zhao13m}. Despite
the qualitative similarity of our results with the helioseismology
inversions, there are some contradictions. The precise determination
of the solar meridional circulation profile remains a matter of the
theoretical and observational progress. 

{The propagation direction of the dynamo wave has been a major
problem plaguing distributed turbulent dynamos since helioseismology
revealed the solar rotation profile \citep{choud95,b05,brsu05}. The
issue is further sharpen by uncertainty of the meridional circulation
distribution inside the solar convection zone. There is a common misconception
in the dynamo community that turbulent dynamos cannot reproduce solar-like
activity. Results of Subsection 3.1 show that it is possible to construct
the distributed dynamo model with the solar-type magnetic cycles.
This requires some tuning of the turbulent pumping coefficients. The
required magnitude of the equatorward pumping velocity is about 3
m/s. Also, in the subsurface layer the negative gradient of the angular
velocity and the positive sign of the $\alpha$-effect provide the
equatorward diffusive drift of the dynamo wave in following to the
Parker-Yoshimura law \citep{1955ApJ...121..491P,yosh1975}. In addition,
the nonlinear models show the increase of the $\alpha$-effect ahead
of the dynamo wave, which amplifies the equatorward drift of the toroidal
magnetic field, see, the second column of Figures \ref{fig:M3} and
\ref{fig:M3a4}. It was found that the resulted pumping velocity distribution
and the effective velocity drift of the large-scale magnetic field
are similar to the recent results of direct numerical simulations
by \citet{2018AA609A..51W}. For the given choice of the turbulent
pumping magnitude, the simulated dynamo wave pattern remains robust
in a range of the surface meridional flow variations $\pm3.5$m/s. }

The dynamo model, which is constructed Subsection 3.1, is applied
to for the numerical study the effect of the magnetic field on the
global flows and heat transport in the solar convection zone. The
large-scale magnetic field affects the global flow by means of both
the mechanical and thermal effects \citep{2006ApJ...647..662R}, who
showed that, in the presence of the meridional circulation, the ``mechanical''
perturbations of the angular momentum transport (associated with the
large- or small-scale Lorentz force) result in the axial distribution
of the torsional oscillation's magnitude inside the convection zone.
This is a consequence of the Taylor-Proudman balance. The results
of the helioseismology show deviations of the torsional oscillation
from the axial distribution. This was interpreted as results magneto-thermal
perturbations. This fact was confirmed in our results. The distribution
of magnitude of the zonal flow variations in the subsurface layer
declines toward the equator in following the dynamo wave. In the model
M3a4 the effect is stronger than in the weakly nonlinear model M3.
The proof requires the further analysis like that made in the paper
by \citet{2016ApJ828L.3G}. This is postponed to another paper.

Here, I would like to stress two facts. Firstly, the numerical model
suggests the pattern of the torsional oscillation which is in qualitative
agreement with observations of \citet{2011JPhCS271a2074H}. Secondly,
the increasing magnetic activity level, for example, due to the $\alpha$
- effect, results in the increasing of the torsional oscillation magnitude
and complication of the meridional circulation structure. Comparing
the model M3 and M3a4, we find that the stagnation point of the bottom
cell is moved to the equator and the upper meridional circulation
cell is divided into two cells. The two stagnation points of the upper
meridional circulation cell result in to two maxima of the meridional
circulation at the surface. There is a similarity between our results
and the recent results of \citet{2017ApJ849.144C} who also found
that the upper circulation cell consists of two. Our model suggest
that it may be a result of the strong toroidal magnetic field in the
subsurface shear. At the surface, the variations of the meridional
circulation in the magnetic cycle show the effective periodical flow
towards the maximum of the toroidal magnetic field. In the model M3
the magnitude of this flow is about $\pm0.5$m/s. The increasing magnetic
activity level in the model M3a4 result in the magnitude of the effective
flow the order of $\pm3.5$m/s and the separation into the equatorial
and polar meridional circulation cells. A similar effect was also
seen in results of \citet{1992AA...265..328B}. Their model has a
different background differential rotation and meridional circulation
(one-cell structure) distributions. Note that in the subsurface layer
the model M3a4 operates at the sub-equipartition level, $\beta\sim0.6-1$,
which can be confronted with Figure \ref{fig:alp}b.

The dynamo model shows the decreasing dynamo period with the increasing
$\alpha$-effect and the increasing level of the magnetic activity.
This is in agreement with results of \citet{pk11} and \citet{pipea2012AA}.
Observations, also, show that, in general, the high solar cycle is
shorter than the low solar cycle \citep{soon93,hath09}. Also, the
shape of the high cycle is different in compare with the low cycles.
This can be deduced in comparison of the time-latitude diagrams of
the model M3 and M3a4, for example, in confronting Figures \ref{fig:m3h}a
and \ref{fig:m34}a, we see that in the model M3a4 the growing phase
of the dynamo wave is much shorter than the decaying phase. In the
model M3, the given asymmetry is less than in the model M3a4. For
the first time, we find these relationships (period-amplitude and
shape-amplitude) in the mean-field model with the meridional circulation.
The flux-transport models fail to explain the relationship between
the solar cycle period and cycle amplitude. In particular, the kinematic
models of \cite{2014ApJ...791...59K} show the increasing dynamo
period with the increasing level of the magnetic activity. This contradicts
to solar and stellar observations (see, e.g., \cite{2009AA_strassm,2017PhDT3E}).
On another hand, the distributed dynamo models are well fitted both
in the solar and stellar observations. Our current results show that
including the meridional circulation in the model does not necessarily
result in the increasing magnetic cycle period with the increasing
level of magnetic field strength. Therefore the issue of the flux-transport
scenario is likely connected with the localized character of the magnetic
field generation effects in the dynamo model.

{For the first time we demonstrate that the Grand activity
cycles can result from effect of the meridional circulation and the
equatorial parity breaking because of the hemispheric magnetic helicity
exchange. The solar data show that hemispheric magnetic helicity exchange
is a real phenomenon \citep{2000JGR...10510481B,2010ApJ719.1955Z,2012ApJ...758...61Y}.
In our model, the process is controlled by diffusion of the magnetic
helicity density and by the magnetic helicity transport with the meridional
circulation.} \citet{mitra10} found this phenomenon in the direct
numerical simulations. In the weakly nonlinear regimes, the strength
of the magnetic helicity diffusion affects the parity of solution.
We find that the symmetric about equator toroidal magnetic field is
generated when the diffusion coefficient is high enough. This is consistent
with the observed increase of the $\alpha$-effect ahead of the dynamo
wave. {The Grand activity cycle regime is not robust and it
disappears when $C_{\alpha}>3C_{\alpha}^{(cr)}$}. This coincides
with the formation of the second meridional circulation cell near
the equator. Currently, it is not clear if both phenomena are tightly
related or this is an accident. This will be studied further.

We show the first results about effects of the large-scale magnetic
activity on the heat transport and the heat energy flux from the dynamo
region. This was previously discussed in papers of \citet{1992AA...265..328B}
and \citet{2000ARep...44..771P}. In the mean-field framework, the
major contributions of the large-scale magnetic field on the heat
energy balance inside the convection zone are caused by the magnetic
quenching of the eddy heat conductivity and the energy expenses (associating
with the heat energy loss and gain) on the large-scale dynamo. These
processes are modeled by the mean-field heat transport equations.
The magnetic perturbations of the heat flux in the model M3 are an
order of $10^{-3}$ of the background value. It is an order of magnitude
less at the surface because of the screening effect and the smaller
strength of the large-scale magnetic field in the upper layer of the
convection zone. The heat perturbation screening effect is due to
the huge heat capacity of the solar convection zone \citep{stix:02}.
Results of the model M3a4 illustrate it better than the model M3.
The model M3a4 shows the strong toroidal magnetic field in the bulk
of the convection zone (see Figure \ref{fig:m34}b). In the upper
layer of the convection zone, the strength of the toroidal field exceeds
the equipartition level. Besides this, the heat flux perturbations
are efficiently smoothed out toward the top of the dynamo region.
Another interesting feature is that the weakly nonlinear model M3
shows the increasing mean heat flux at the maximum of the magnetic
cycle. In the model with the overcritical $\alpha$ effect, we find
the opposite situation. The solar observations show the increasing
luminosity during the maximum of the solar cycles \citep{1999GeoRL26.3613W}.
The variation of the photometric brightness of solar-type stars tends
to inverse the sign with the increasing level of the magnetic activity
\citep{2016ASPC504.273Y}. From the point of view of our model, this
means that the effect of the magnetic shadow become dominant when
the total magnetic activity is increased. This is a preliminary conclusion.
Also, the relationship between the magnetic shadow effect in the large-scale
dynamo and the stellar surface darkening because of starspots is not
straightforward.

Finally, our results can be summarized as follows: 
\begin{enumerate}
\item We constructed the nonkinematic solar-type dynamo model with the double-cell
meridional circulation. The role of the turbulent pumping in the dynamo
model should be investigated. This requires a better theoretical and
observational knowledge of the solar meridional circulation. 
\item The torsional oscillations are explained as a result of the magnetic
feedback on the angular momentum transport by the turbulent stresses,
the effect of the Lorentz force and the magneto-thermal perturbations
of the Taylor-Proudman balance. The increasing level of the magnetic
activity results in separation of the upper meridional circulation
cell for two parts. 
\item The model shows the decrease of the dynamo period with the increase
of the magnetic cycle amplitude. The shape of the strong magnetic
cycle is more asymmetric than the shape of the weak cycles. 
\item The magnetic helicity density diffusion and the increase turbulent
generation of the large-scale magnetic field results in the increasing
hemispheric magnetic helicity exchange, the magnetic parity breaking
and the Grand activity cycles. The Grand activity cycles exists in
the intermediate range of the $\alpha$ - effect parameter, when $1.5C_{\alpha}^{(cr)}<C_{\alpha}<3C_{\alpha}^{(cr)}$.
It seems that the Grand activity cycles disappear together with the
formation of the second meridional circulation cell near the equator. 
\item The increasing turbulent generation of the large-scale magnetic field
changes of the relationship between the magnetic cycle phase and the
mean sign of the heat flux perturbation at the surface. For the high
level of the magnetic activity, the heat flux is reduced in the maximum
of the magnetic cycle. 
\end{enumerate}
{Acknowledgments.} This work was conducted as a part of FR
II.16 of ISTP SB RAS. Author thanks the financial support from of
RFBR grant 17-52-53203. 
\begin{center}
{REFERENCES} 
\par\end{center}

\bibliographystyle{elsarticle-harv}

\section*{Appendix}

\subsection*{Heat transport\label{subsec:Heat-transport}}

\citet{phd} found that under the joint action of the Coriolis force
and the large-scale\emph{ toroidal} magnetic field, and when it holds
$\Omega^{*}>1$, the eddy heat conductivity tensor could be approximated
as follows 
\begin{equation}
\chi_{ij}\approx\chi_{T}\left(\phi_{\chi}^{(I)}\left(\beta\right)\phi\left(\Omega^{*}\right)\delta_{ij}+\phi_{\chi}^{(\|)}\left(\beta\right)\phi_{\parallel}\left(\Omega^{*}\right)\frac{\Omega_{i}\Omega_{j}}{\Omega^{2}}\right),\label{eq:ht-F-1}
\end{equation}
where functions $\phi$ and $\phi_{\parallel}$ were defined in \citet{1994AN....315..157K},
and the magnetic quenching functions $\phi_{\chi}^{(I)}$ and $\phi_{\chi}^{(\|)}$
are 
\begin{eqnarray*}
\phi_{\chi}^{(I)} & = & \frac{2}{\beta^{2}}\left(1-\frac{1}{\sqrt{1+\beta^{2}}}\right),\\
\phi_{\chi}^{(\|)} & = & \frac{2}{\beta^{2}}\left(\sqrt{1+\beta^{2}}-1\right).
\end{eqnarray*}
where $\beta=\left|\overline{\mathbf{B}}\right|/\sqrt{4\pi\overline{\rho}u'^{2}}$.
The difference of the Eq(\ref{eq:ht-F-1}) from results of \citet{1994AN....315..157K}
is that for $\Omega^{*}\gg1$ and $\beta\gg1$ the isotropic and anisotropic
part of the eddy heat conductivity tensor become close.

\subsection*{The turbulent stress tensor}

Expression of the turbulent stress tensor results from the mean-field
hydrodynamics theory (see, \cite{1994AN....315..157K,kit2004AR})
as follows 
\begin{equation}
{\hat{T}_{ij}=\left(\left\langle u_{i}u_{j}\right\rangle -\frac{1}{4\pi\overline{\rho}}\left(\left\langle b_{i}b_{j}\right\rangle -\frac{1}{2}\delta_{ij}\left\langle \mathbf{b}^{2}\right\rangle \right)\right),}\label{eq:stres-1}
\end{equation}
where $\mathbf{u}$ and $\mathbf{b}$ are fluctuating velocity and
magnetic fields. Application the mean-field hydrodynamic framework
leads to the Taylor expansion: 
\begin{eqnarray}
\hat{T}_{ij} & = & \hat{T}_{ij}^{\left(\Lambda\right)}+\hat{T}_{ij}^{\left(\nu\right)}\label{eq:tstr}\\
 & = & \Lambda_{ijk}\Omega_{k}-N_{ijkl}\frac{\partial\overline{U}_{k}}{\partial r_{l}}+\dots
\end{eqnarray}
where the first term represent turbulent generation of the large-scale
flow and the second one stands for the dissipative effect. The viscous
part of the azimuthal components of the stress tensor is determined
following to \citet{1994AN....315..157K} in this form: 
\begin{eqnarray}
{T_{r\phi}^{(\nu)}} & {=} & -{\nu_{T}\left\{ \Phi_{\perp}+\left(\Phi_{\|}-\Phi_{\perp}\right)\mu^{2}\right\} r\frac{\partial\sin\theta\Omega}{\partial r}}\label{eq:trf}\\
 & - & {\nu_{T}\sin\theta\left(\Phi_{\|}-\Phi_{\perp}\right)\left(1-\mu^{2}\right)\frac{\partial\Omega}{\partial\mu}}\nonumber \\
{T_{\theta\phi}^{(\nu)}} & {=} & {\nu_{T}\sin^{2}\theta\left\{ \Phi_{\perp}+\left(\Phi_{\|}-\Phi_{\perp}\right)\sin^{2}\theta\right\} \frac{\partial\Omega}{\partial\mu}}\label{eq:ttf}\\
 & + & {\nu_{T}\left(\Phi_{\|}-\Phi_{\perp}\right)\mu\sin^{2}\theta r\frac{\partial\Omega}{\partial r}},\nonumber 
\end{eqnarray}
where the eddy viscosity, $\nu_{T}$, is determined from the mixing-length
theory assuming the turbulent Prandtl number ${Pr}_{T}={\displaystyle \frac{3}{4}}$:

\[
\nu_{T}=\frac{3\ell^{2}}{16}\sqrt{-\frac{g}{2c_{p}}\frac{\partial\overline{{s}}}{\partial r}}.
\]
The viscosity quenching functions, $\Phi_{\|}$ and $\Phi_{\perp}$,
depend nonlinearly on the Coriolis number, $\Omega^{*}=2\Omega\tau_{c}$,
and the strength of the large-scale magnetic field. In the model we
employ the analytic expressions for the magnetic quenching functions
of the eddy viscosity and the the $\Lambda$- effect obtained by \citet{1994AN....315..157K},
\citet{kuetal96} and \citet{p99} for the fast rotating regime ($\Omega^{*}>1$):
\begin{eqnarray}
{\Phi_{\perp}} & {=} & \psi_{\perp}\left(\Omega^{\star}\right)\left(\phi_{V\perp}\left(\beta\right)+\phi_{\chi}^{(I)}\left(\beta\right)\right)\label{visc-f}\\
\Phi_{\parallel} & = & {\psi_{\parallel}\left(\Omega^{\star}\right)\phi_{\chi}^{(I)}\left(\beta\right),}
\end{eqnarray}
where the $\psi_{\perp}$ and $\psi_{\parallel}$ are determined by
\citet{1994AN....315..157K} and: 
\begin{eqnarray}
{\phi_{\chi}^{(I)}} & {=} & {\frac{2}{\beta^{2}}\left(1-\frac{1}{\sqrt{1+\beta^{2}}}\right),}\\
{\phi_{V\perp}} & {=} & {\frac{4}{\beta^{4}\sqrt{\left(1+\beta^{2}\right)^{3}}}\left(\left(\beta^{4}+19\beta^{2}+18\right)\sqrt{\left(1+\beta^{2}\right)}\right.}\\
 & - & \left.8\beta^{4}-28\beta^{2}-18\right).\nonumber 
\end{eqnarray}

The non-diffusive flux of angular momentum $\boldsymbol{\Lambda}=\left\langle {u}_{\phi}'{\boldsymbol{u}}\right\rangle $
can be expressed as follows \citep{1989drsc.book.....R}: 
\begin{eqnarray}
\hat{T}_{r\phi}^{\left(\Lambda\right)} & = & r\Lambda_{V}\Omega\sin\theta,\nonumber \\
\Lambda_{V} & = & \nu_{T}\left(V^{(0)}+\sin^{2}\theta V^{(1)}\right),\label{eq:lv}\\
\hat{T}_{\theta\phi}^{\left(\Lambda\right)} & = & r\Lambda_{H}\Omega\cos\theta,\nonumber \\
\Lambda_{H} & = & \nu_{T}\left(H^{(0)}+\sin^{2}\theta H^{(1)}\right)\label{eq:lh}
\end{eqnarray}
The basic contributions to the $\Lambda$-effect are due to the density
stratification and the Coriolis force \citep{KR93L}. The analytical
form of the $\Lambda$-effect coefficients become fairly complicated
if we wish to account the multiple-cell meridional circulation structure.
In particular, Pipin \& Kosovichev (2017) found that the spatial derivative
of the Coriolis number $\Omega^{*}=2\Omega_{0}\tau_{c}$, has to be
taken into accounted. In nonlinear model we take into account the
effect of magnetic field (see, \citep{p99}). Also effect of the convective
velocities anisotropy is important to account the subsurface shear
layer (see, \citep{kit2004AR}). Therefore, the final coefficients
of the $\Lambda$-tensor are: 
\begin{eqnarray}
V^{(0)} & = & \Bigl[\left(\frac{\alpha_{MLT}}{\gamma}\right)^{2}\left\{ J_{0}+\!J_{1}+\!a\left(I_{0}+\!I_{1}\right)\right\} \label{v0-f}\\
 & - & \Bigl(\frac{\alpha_{MLT}\ell}{\gamma}\frac{\partial}{\partial r}\left\{ \left(J_{0}+J_{1}\right)-I_{5}+I_{6}\right\} +\ell^{2}\frac{\partial^{2}}{\partial r^{2}}\left(I_{1}-I_{2}\right)\Bigr]\phi_{\chi}^{(I)}\left(\beta\right),\nonumber \\
V^{(1)} & = & -\left\{ \left(\frac{\alpha_{MLT}}{\gamma}\right)^{2}\left(J_{1}+aI_{1}\right)-\frac{\alpha_{MLT}\ell}{\gamma}\frac{\partial}{\partial r}\left(J_{1}+I_{6}\right)-\ell^{2}\frac{\partial^{2}}{\partial r^{2}}I_{2}\right\} \phi_{\chi}^{(I)}\left(\beta\right),\label{v1-f}
\end{eqnarray}
and ${H^{(1)}=-V^{(1)}}$. We employ the parameter of the turbulence
anisotropy $a={\displaystyle \frac{\overline{u_{h}^{2}}-2\overline{u_{r}^{2}}}{\overline{u_{r}^{2}}}=}2$,
where $u_{h}$ and $u_{r}$ are the horizontal and vertical RMS velocities
\citep{kit2004AR}. Collecting results of \citet{kit-rud-kuk} and
\citet{kuetal96} we write the coefficient $H^{(0)}$ as follows:
\begin{equation}
H^{(0)}=\frac{\tau^{2}}{\rho^{2}}J_{4}\phi_{H}\left(\beta\right)\frac{\partial^{2}}{\partial r^{2}}\left\langle \boldsymbol{u}'^{2}\right\rangle \rho^{2}=\left\{ 4\left(\frac{\alpha_{MLT}}{\gamma}\right)^{2}J_{4}-\frac{\alpha_{MLT}\ell}{\gamma}\frac{\partial}{\partial r}J_{4}-\ell^{2}\frac{\partial^{2}}{\partial r^{2}}J_{4}\right\} \phi_{H}\left(\beta\right),\label{h0-f}
\end{equation}
where function $J_{4}$ was defined in \citet{kit-rud-kuk} and the
magnetic quenching function $\phi_{H}\left(\beta\right)$ was defined
by \citet{p99}: 
\begin{eqnarray}
\phi_{H} & = & \frac{1}{\beta^{2}}\left(\frac{2+3\beta^{2}}{2\sqrt{\left(1+\beta^{2}\right)^{3}}}-1\right).
\end{eqnarray}
Note that $\phi_{H}\approx-\beta^{2}$ for the small magnetic field
strength. Therefore the coefficient $H^{(0)}$ disappears in the absence
of the large-scale magnetic field. The quenching functions of the
the Coriolis number $\Omega^{*}$, $J_{n}$and $I_{n}$were defined
in \citet{KR93L} and \citet{kit-rud-kuk}. For convenience, they
are given below. Functions $I_{n}$ have the following form: 
\begin{eqnarray*}
I_{1} & = & \frac{1}{4\Omega^{*4}}\left(\frac{6+5\Omega^{*2}}{1+\Omega^{*2}}-\left(6+\Omega^{*2}\right)\frac{\arctan\Omega^{*}}{\Omega^{*}}\right),\\
I_{2} & = & \frac{1}{8\Omega^{*4}}\Bigl(60+\Omega^{*2}-\frac{6\Omega^{*2}}{1+\Omega^{*2}}\\
 & + & \left(\Omega^{*4}-15\Omega^{*2}-60\right)\frac{\arctan\Omega^{*}}{\Omega^{*}}\Bigr),\\
I_{3} & = & \frac{1}{2\Omega^{*4}}\left(-3+\frac{\Omega^{*2}}{1+\Omega^{*2}}+3\frac{\arctan\Omega^{*}}{\Omega^{*}}\right),\\
I_{4} & = & \frac{1}{2\Omega^{*4}}\left(-15+\frac{2\Omega^{*2}}{1+\Omega^{*2}}+\left(15+3\Omega^{*2}\right)\frac{\arctan\Omega^{*}}{\Omega^{*}}\right),\\
I_{5} & = & \frac{1}{2\Omega^{*4}}\left(-3+\left(\Omega^{*2}+3\right)\frac{\arctan\Omega^{*}}{\Omega^{*}}\right),\,I_{6}=\frac{1}{2}I_{4},\\
J_{4} & = & \frac{1}{16\Omega^{*6}}\Bigl(40-\frac{31}{3}\Omega^{*2}-3\Omega^{*4}\\
 & - & \left(3\Omega^{*6}-10\Omega^{*4}+3\Omega^{*2}+40\right)\frac{\arctan\Omega^{*}}{\Omega^{*}}\Bigr)
\end{eqnarray*}
In addition, we have $J_{0}=4I_{1}+2I_{5}$ and $J_{1}=-4I_{2}-2I_{6}$,
see the above cited paper.

The first RHS term of Eq.(\ref{eq:vort}) describes dissipation of
the mean vorticity, $\omega$. It has a combersome expression (see,\citealp{1989drsc.book.....R}):
\begin{eqnarray}
-\left[\boldsymbol{\nabla}\times\frac{1}{\overline{\rho}}\boldsymbol{\nabla\cdot}\overline{\rho}\hat{\mathbf{T}}\right]_{\phi} & = & \frac{\sin\theta}{r^{2}}\frac{\partial^{2}}{\partial\mu^{2}}\sin\theta\psi_{1}\left(\beta\right)\hat{T}_{r\theta}^{\left(\nu\right)}-\frac{1}{r}\frac{\partial}{\partial r}\frac{1}{r^{2}\overline{\rho}}\frac{\partial}{\partial r}r^{3}\overline{\rho}\psi_{1}\left(\beta\right)\hat{T}_{r\theta}^{\left(\nu\right)}\label{eq:vort2-1}\\
 & - & \frac{1}{r}\frac{\partial}{\partial r}\frac{\partial}{\partial\mu}\sin\theta\psi_{1}\left(\beta\right)\left(\hat{T}_{rr}^{\left(\nu\right)}-\hat{T}_{\theta\theta}^{\left(\nu\right)}\right)+\frac{\cot\theta}{r}\frac{\partial}{\partial r}\psi_{1}\left(\beta\right)\left(\hat{T}_{\phi\phi}^{\left(\nu\right)}-\hat{T}_{rr}^{\left(\nu\right)}\right)\\
 & + & \frac{\sin\theta}{r^{2}}\frac{\partial}{\partial\mu}\psi_{1}\left(\beta\right)\left(\hat{T}_{\theta\theta}^{\left(\nu\right)}+\hat{T}_{\phi\phi}^{\left(\nu\right)}-2\hat{T}_{rr}^{\left(\nu\right)}\right)-\frac{\sin\theta}{r\overline{\rho}}\frac{\partial\overline{\rho}}{\partial r}\frac{\partial}{\partial\mu}\psi_{1}\left(\beta\right)\hat{T}_{rr}^{\left(\nu\right)}
\end{eqnarray}
where, $\mu=\cos\theta$, the components of $\hat{T}_{ij}^{\left(\nu\right)}$
are given in \citet{1994AN....315..157K}, the magnetic quenching
function $\psi_{1}\left(\beta\right)$ (see the above cited paper)
takes into account the magnetic feedback on the eddy-viscosity tensor
in the simplest way.

\subsection*{The $\alpha$ and $\eta$ tensors}

The $\alpha$- effect takes into account the kinetic and magnetic
helicities, 
\begin{eqnarray}
\alpha_{ij} & = & C_{\alpha}\psi_{\alpha}(\beta)\alpha_{ij}^{(H)}\eta_{T}+\alpha_{ij}^{(M)}\frac{\overline{\chi}\tau_{c}}{4\pi\overline{\rho}\ell^{2}}\label{alp2d-1}
\end{eqnarray}
where $C_{\alpha}$ is a free parameter, the $\alpha_{ij}^{(H)}$
and $\alpha_{ij}^{(M)}$ express the kinetic and magnetic helicity
coefficients, respectively, $\overline{\chi}$- is the small-scale
magnetic helicity, and $\ell$ is the typical length scale of the
turbulence. The helicity coefficients have been derived by \citet{pi08Gafd}
(hereafter P08). The $\alpha_{ij}^{(H)}$ reads, 
\begin{eqnarray}
\alpha_{ij}^{(H)} & = & \delta_{ij}\left\{ 3\left(f_{10}^{(a)}\left(\mathbf{e}\cdot\boldsymbol{\Lambda}^{(\rho)}\right)+f_{11}^{(a)}\left(\mathbf{e}\cdot\boldsymbol{\Lambda}^{(u)}\right)\right)\right\} +\label{eq:alpha}\\
 & + & e_{i}e_{j}\left\{ 3\left(f_{5}^{(a)}\left(\mathbf{e}\cdot\boldsymbol{\Lambda}^{(\rho)}\right)+f_{4}^{(a)}\left(\mathbf{e}\cdot\boldsymbol{\Lambda}^{(u)}\right)\right)\right\} \nonumber \\
 & + & 3\left\{ \left(e_{i}\Lambda_{j}^{(\rho)}+e_{j}\Lambda_{i}^{(\rho)}\right)f_{6}^{(a)}+\left(e_{i}\Lambda_{j}^{(u)}+e_{j}\Lambda_{i}^{(u)}\right)f_{8}^{(a)}\right\} ,\nonumber 
\end{eqnarray}
where $\mathbf{e}={\displaystyle \frac{\boldsymbol{\Omega}}{\Omega}},$
$\mathbf{\boldsymbol{\Lambda}}^{(\rho)}=\boldsymbol{\nabla}\log\overline{\rho}$
, $\mathbf{\boldsymbol{\Lambda}}^{(u)}=\boldsymbol{\nabla}\log\left({u'}\ell\right)$
and the $\alpha_{ij}^{(M)}$ reads: 
\begin{equation}
\alpha_{ij}^{(M)}=2f_{2}^{(a)}\delta_{ij}-2f_{1}^{(a)}e_{i}e_{j},\label{alpM}
\end{equation}
Functions $f_{n}^{(a)}\left(\Omega^{*}\right)$ were defined by P08.
The magnetic quenching function of the hydrodynamical part of $\alpha$-effect
is defined by 
\begin{equation}
\psi_{\alpha}=\frac{5}{128\beta^{4}}\left(16\beta^{2}-3-3\left(4\beta^{2}-1\right)\frac{\arctan\left(2\beta\right)}{2\beta}\right).
\end{equation}
In the notations of P08 :$\psi_{\alpha}=-3/4\phi_{6}^{(a)}$. The
dependence of the $\alpha_{ij}^{(H)}$ and $\alpha_{ij}^{(M)}$ tensors
on the Coriolis number is defined in following P08: 
\begin{eqnarray*}
f_{1}^{(a)} & = & \frac{1}{4\Omega^{*\,2}}\left(\left(\Omega^{*\,2}+3\right)\frac{\arctan\Omega^{*}}{\Omega^{*}}-3\right),\\
f_{2}^{(a)} & = & \frac{1}{4\Omega^{*\,2}}\left(\left(\Omega^{*\,2}+1\right)\frac{\arctan\Omega^{*}}{\Omega^{*}}-1\right),\\
f_{4}^{(a)} & = & \frac{1}{6\Omega^{*\,3}}\left(3\left(\Omega^{*4}+6\varepsilon\Omega^{*2}+10\varepsilon-5\right)\frac{\arctan\Omega^{*}}{\Omega^{*}}\right.\\
 & - & \left.\left((8\varepsilon+5)\Omega^{*2}+30\varepsilon-15\right)\right),\\
f_{5}^{(a)} & = & \frac{1}{3\Omega^{*\,3}}\left(3\left(\Omega^{*4}+3\varepsilon\Omega^{*2}+5(\varepsilon-1)\right)\frac{\arctan\Omega^{*}}{\Omega^{*}}\right.\\
 & - & \left.\left((4\varepsilon+5)\Omega^{*2}+15(\varepsilon-1)\right)\right),\\
f_{6}^{(a)} & = & -\frac{1}{48\Omega^{*\,3}}\left(3\left(\left(3\varepsilon-11\right)\Omega^{*2}+5\varepsilon-21\right)\frac{\arctan\Omega^{*}}{\Omega^{*}}\right.\\
 & - & \left.\left(4\left(\varepsilon-3\right)\Omega^{*2}+15\varepsilon-63\right)\right),\\
f_{8}^{(a)} & = & -\frac{1}{12\Omega^{*\,3}}\left(3\left(\left(3\varepsilon+1\right)\Omega^{*2}+4\varepsilon-2\right)\frac{\arctan\Omega^{*}}{\Omega^{*}}\right.\\
 & - & \left.\left(5\left(\varepsilon+1\right)\Omega^{*2}+12\varepsilon-6\right)\right),\\
f_{10}^{(a)} & = & -\frac{1}{3\Omega^{*\,3}}\left(3\left(\Omega^{*2}+1\right)\left(\Omega^{*2}+\varepsilon-1\right)\frac{\arctan\Omega^{*}}{\Omega^{*}}\right.\\
 & - & \left.\left(\left(2\varepsilon+1\right)\Omega^{*2}+3\varepsilon-3\right)\right),\\
f_{11}^{(a)} & = & -\frac{1}{6\Omega^{*\,3}}\left(3\left(\Omega^{*2}+1\right)\left(\Omega^{*2}+2\varepsilon-1\right)\frac{\arctan\Omega^{*}}{\Omega^{*}}\right.\\
 & - & \left.\left(\left(4\varepsilon+1\right)\Omega^{*2}+6\varepsilon-3\right)\right).
\end{eqnarray*}
Note, that the parameter $\varepsilon={\displaystyle \frac{\overline{\mathbf{b}^{2}}}{\mu_{0}\overline{\rho}\overline{\mathbf{u}^{2}}}}$,
control the theoretical ratio between the magnetic and kinetic energies
of fluctuations in the background turbulence. It is assumed that $\varepsilon=1$.

We employ the anisotropic diffusion tensor which is derived in P08
and in \citep{2014ApJ_pipk}: 
\begin{eqnarray}
\eta_{ijk} & = & 3\eta_{T}\left\{ \left(2f_{1}^{(a)}-f_{2}^{(d)}\right)\varepsilon_{ijk}+2f_{1}^{(a)}\frac{\Omega_{i}\Omega_{n}}{\Omega^{2}}\varepsilon_{jnk}\right\} \label{eq:diff}\\
 & + & a\eta_{T}\phi_{1}\left(g_{n}g_{j}\varepsilon_{ink}-\varepsilon_{ijk}\right)\nonumber 
\end{eqnarray}
where $\mathbf{g}$ is the unit vector in the radial direction. We
employ the same turbulence anisotropy parameter $a=2$ as for the
$\Lambda$ effect. The quenching functions $f_{2}^{(d)}$and $\phi_{1}$
are determined in P08 and in \citet{2014ApJ_pipk}:
\begin{eqnarray}
f_{2}^{(d)} & = & \frac{1}{2\Omega^{*\,3}}\left(\left(\varepsilon+1\right)\Omega^{*\,2}+3\varepsilon\right.\label{eq:phi1}\\
 & - & \left.\left(\left(2\varepsilon+1\right)\Omega^{*\,2}+3\varepsilon\right)\frac{\arctan\left(\Omega^{*}\right)}{\Omega^{*}}\right)\\
\phi_{1} & = & -\frac{1}{24\Omega^{\star2}}\left(2\log\left(1+4\Omega^{\star2}\right)+4\log\left(1+\Omega^{\star2}\right)+\right.\\
 &  & +\left.\left(1-4\Omega^{\star2}\right)\frac{\arctan\left(2\Omega^{\star}\right)}{\Omega^{\star}}+4\left(1-\Omega^{\star2}\right)\frac{\arctan\left(\Omega^{\star}\right)}{\Omega^{\star}}-6\right).\nonumber 
\end{eqnarray}

\end{document}